\documentclass[useAMS,usenatbib]{mn2e}
\usepackage{graphicx}
\usepackage{amsmath}
\usepackage{bm}

\newcommand{\fig}[1]{\mbox{Fig.~\ref{#1}}}

\newcommand{\eq}[1]{\mbox{equation~(\ref{#1})}}

\newcommand{\omegat}{\mbox{$\omega\left(\theta\right)$}}
\newcommand{\omegals}{\mbox{$\omega_{\rm LS}$}}

\newcommand{\ic}{\mbox{$\mathcal{I}$}}

\newcommand{\tp}{\mbox{$\left(\theta\right)$}}

\newcommand{\DD}{\mbox{$DD$}}
\newcommand{\RR}{\mbox{$RR$}}
\newcommand{\DR}{\mbox{$DR$}}

\newcommand{\BzK}{\mbox{$BzK$}}
\newcommand{\bzk}{$BzK$}
\newcommand{\gzHK}{\mbox{$gzHK$}}

\newcommand{\sgzHK}{\mbox{SF-$gzHK$}}
\newcommand{\pgzHK}{\mbox{PE-$gzHK$}}
\newcommand{\gzhk}{$gzHK$}
\newcommand{\gzHKs}{$gzHK$}

\newcommand{\Aw}{\mbox{$A_\omega$}}
\newcommand{\g}{$\gamma$}

\newcommand{\Msun}{\mbox{$M_\odot$}}

\newcommand{\z}{$z$}
\newcommand{\zs}{$z$$\sim$}
\newcommand{\chisq}{$\chi^2$}

\newcommand{\aap}{\mbox{A\&A}}

\newcommand{\aj}{\mbox{AJ}}
\newcommand{\apj}{\mbox{ApJ}}
\newcommand{\apjl}{\mbox{ApJL}}
\newcommand{\apjs}{\mbox{ApJS}}
\newcommand{\araa}{\mbox{ARA\&A}}
\newcommand{\mnras}{\mbox{MNRAS}}

\title[Clustering of z$\sim$2 galaxies]{
Angular clustering of $z$$\sim$2 star-forming and passive galaxies in 2.5 square degrees of deep CFHT imaging
}

\author[T.\ Sato, M.\ Sawicki, and L.\ Arcila-Osejo]{
Taro Sato,
Marcin Sawicki\thanks{Corresponding author. E-mail: sawicki@ap.smu.ca.},
	  and 
  Liz Arcila-Osejo
    \\ Department of Astronomy \& Physics and the Institute for Computational Astrophysics, Saint Mary's University, \\ 923 Robie Street, Halifax, Nova Scotia, B3H 3C3, Canada}
\begin{document}

\date{Accepted for publication in MNRAS}

\pagerange{\pageref{firstpage}--\pageref{lastpage}} \pubyear{2002}

\maketitle

\label{firstpage}

\begin{abstract}

We study the angular clustering of $z \sim 2$ galaxies using $\sim$40,000 star-forming (SF) and $\sim$5,000 passively-evolving (PE) galaxies selected from $\sim$2.5 deg$^2$ of deep ($K_{lim}$=23--24 AB) CFHT imaging. For both populations the clustering is stronger for galaxies brighter in rest-frame optical and the trend is particularly strong for PE galaxies, indicating that passive galaxies with larger stellar masses reside in more massive halos. In contrast, at rest-frame UV we find that while the clustering of SF galaxies increases with increasing luminosity, it decreases for PE galaxies; a possible explanation lies in quenching of star formation in the most massive halos. Furthermore, we find two components in the correlation functions for both SF and PE galaxies, attributable to one- and two-halo terms. The presence of one-halo terms for both PE and SF galaxies suggests that environmental effects were producing passive galaxies in virtualized environments already by $ z \sim 2$.  Finally, we find notable clustering differences  between the four widely-separated fields in our study; the popular COSMOS field is the most discrepant (as is also the case for number counts and luminosity functions), highlighting the need for very large areas and multiple sightlines in galaxy evolution statistical studies. 

\end{abstract}


\section{Introduction}\label{sec.1}

A wide array of evidence has led to a general agreement that the lambda cold dark matter model ($\Lambda$CDM) successfully describes the structure evolution driven by matter-energy distribution, much of which is dominated by observationally inaccessible dark components (e.g., Komatsu et al.\ 2011). What has become vastly more accessible through observation during the last couple decades is the luminous parts of the universe, i.e., galaxies. Through large-scale redshift surveys as SDSS (York et al.\ 2000), 2dFGRS (Colless et al.\ 2001), DEEP2 (Newman et al.\ 2012), and zCOSMOS (Lilly et al.\ 2007), there is now a wide agreement that star-forming activity in the universe peaked at z$\sim$2 (see, e.g., Hopkins \& Beacom 2006, for a compilation) --- by which time roughly a fifth of the present-day stellar mass was already in place (e.g., Sawicki 2012a) --- and the star-forming activity in the universe has been steadily declining ever since (e.g., Lilly et al.\ 1996; Noeske et al.\ 2007). This overall trend in star formation history (SFH) of galaxies has been made most clear in the form of the Madau plot (e.g., Madau et al.\ 1996; Lilly et al.\ 1996; Sawicki et al.\ 1997), showing the luminosity density as a function of redshift. Constraining the physical mechanism that gives rise to this trend in SFH has been one of the most pressing issues in extragalactic astrophysics.

The relative importance of internal versus external processes on the evolution of galaxies has been extensively debated. While the importance of baryonic physics has been realized early in the canonical models of galaxy formation (e.g., Rees \& Ostriker 1977; Silk 1977), the significance of supernovae/AGN feedback in the self-regulation of star-forming activities has been suggested relatively recently (e.g., Dekel \& Silk 1986; Bower et al.\ 2006; Croton et al.\ 2006). The environmental effects on the properties of galaxies were already remarked in much earlier times (e.g., Hubble 1936) and have been well established ever since the seminal work by Dressler (1980), quantifying the famous morphology-density relation in clusters of galaxies. In the local universe, quiescent galaxies are found preferentially in denser environments (e.g., Balogh et al.\ 2004). In the past, the nearest-neighbor approach was popular in defining the environment surrounding galaxies. At high redshifts, however, the method suffers from lack of sufficiently precise redshift measurements, where spectroscopy is expensive and photometric redshifts require a wide and well-sampled multi-wavelength baseline for constraining the spectral energy distributions (SEDs).

Another oft-used approach to quantifying galaxy environment is to construct a two-point correlation function. From theories and simulations, the clustering properties of dark haloes have been well quantified (e.g., Mo \& White 1996). Observationally, the clustering measurement of luminous galaxies is relatively straightforward through imaging data. This makes possible the comparisons between the clustering properties of luminous and dark components of galaxies, often in the form of galaxy bias, the ratio of galaxy to dark halo distributions. Of particular interest is the connection between luminous galaxies and the mass of their dark halo hosts, which broadly defines the host environment of galaxies. For the most massive dark halos (i.e., galaxy clusters), mass estimates may be obtained via X-ray emission, gravitational lensing, cluster-member kinematics, and the Sunyaev-ZelÕdovich effect. For less massive halos, however, similar methods are less accessible due to observational costs. On the other hand, clustering measurement offers an efficient way of quantifying the relation between galaxies and dark halos from ever-increasing catalogs of photometric observations. In fact, the method has been so effective and lead to the development of the halo occupation distribution (HOD) framework (e.g., Jing 1998; Ma \& Fry 2000; Peacock \& Smith 2000; Seljak 2000; Scoccimarro et al.\ 2001; Berlind \& Weinberg 2002; Cooray \& Sheth 2002; Yang et al.\ 2003; Kravtsov et al.\ 2004; Zheng et al.\ 2005) to interpret observations, which describes the probability that a dark halo of virial mass $M_h$ hosts N galaxies of a given set of properties.

In the local universe, the vast amount of observations from large redshift surveys have steadily improved the measurements of galaxy clustering in terms of their intrinsic properties including luminosity, color, morphology, and starforming properties (e.g., Norberg et al.\ 2001, 2002; Zehavi et al.\ 2002, 2005; Budav\'ari et al.\ 2003; Madgwick et al.\ 2003; Li et al.\ 2006; Swanson et al.\ 2008; Ross \& Brunner 2009; Loh et al.\ 2010; Ross et al.\ 2010; Zehavi et al.\ 2011). Their findings are generally consistent with other approaches on galaxy environments, such as nearest-neighbor measurements. The types of galaxies that tend to cluster more strongly (i.e., denser environments) are luminous/massive, bulge-dominated, and redder/quiescent galaxy populations. Such environmental dependence persists to intermediate redshift z$\sim$1 (e.g., Coil et al.\ 2004; Le F\'evre et al.\ 2005; Coil et al.\ 2006; Phleps et al.\ 2006; Pollo et al.\ 2006; Coil et al.\ 2008; Meneux et al.\ 2008; McCracken et al.\ 2008; Meneux et al.\ 2009; Simon et al.\ 2009; Abbas et al.\ 2010).

At $z > 1$, where extensive spectroscopy is difficult, a wide array of photometric color selection techniques have been devised. These rely on prominent ÒbreaksÓ in the spectral energy distribution, such as hydrogen Lyman/Balmer breaks to red-shift in between broadband filters, causing substantial color changes. The Lyman-break dropout technique (Guhathakurta et al.\ 1990; Steidel et al.\ 1996), for example, made it feasible to cull a large number of star-forming $z\sim3$ galaxies purely through photometry. Due to the popularity of the technique, the clustering of Lyman-break galaxies has been well studied (e.g., Adelberger et al.\ 2005; Ouchi et al.\ 2005; Lee et al.\ 2006; Hildebrandt et al.\ 2009; Bielby et al.\ 2012; Savoy et al.\ 2011), showing that clustering strength tends to increase with the UV luminosity of galaxies. However, the galaxies selected in rest-frame UV are biased toward star-forming systems with little dust, and the characterization cannot be extended to wider populations. Various color selection methods in similar spirit have been devised to target $z > 1$ galaxies: EROs (e.g., Elston et al.\ 1988; McCarthy et al.\ 1992; Hu \& Ridgway 1994; Thompson et al.\ 1999; McCarthy 2004), DRGs (Franx et al.\ 2003), and \BzK\ (Daddi et al.\ 2004). These color selection techniques are efficient, requiring only a few photometric bandpasses, but in general select biased populations, making difficult the comparisons between samples selected differently (e.g., Reddy et al.\ 2005; Lane et al.\ 2007; Grazian et al.\ 2007).

The \BzK\ selection technique (Daddi et al.\ 2004) has been well-tested and employed in several studies of $z\sim2$ galaxy clustering (e.g., Kong et al.\ 2006; Hayashi et al.\ 2007; Blanc et al.\ 2008; Hartley et al.\ 2008; McCracken et al.\ 2010; Ly et al.\ 2011; Fang et al.\ 2012; Lin et al.\ 2012). The popularity of the method comes in part from its ability to construct relatively complete samples of star-forming and passively-evolving galaxies at $z\sim2$ without regard for dust reddening. Typically, a magnitude-limited source catalog is constructed from the $K$ band image. While sampling redder light in near-infrared is desirable for tracing stellar mass reliably, the lack of deep, wide-field $K$-band imaging has been a bottleneck for \BzK\ clustering surveys in the past. Furthermore, a clustering measurement from a small field of view suffers from the variance in observation inherent in choosing a particular sightline in the universe (i.e., cosmic variance). A common approach to combat this is to observe a few independent sightlines and/or bootstrap a set of regions within an image to simulate a different set of observations. The wide-field near-infrared imagers (e.g., KPNO/NEWFIRM and CFHT/WIRCam) of the current generation, however, have been making data accessible lately to lessen these issues.

The visible/near-infrared imaging data from the Canada-France-Hawaii Telescope (CFHT) Legacy Survey offer desirable features for clustering measurements. In visible, the CFHT/MegaPrime imaging consists of four independent fields, each being a 1 square degree image, widely separated from each other on the sky. The near-infrared imaging data from the CFHT/WIRCam are mosaics of 21.5$\arcmin$ square images, which makes the effective survey areas smaller, but are comparably deep as the current generation of $K$-limited surveys. These data allow us to study the clustering properties of $z\sim2$ galaxies in unprecedented scale in terms of number of objects and the survey area. The original \BzK\ technique by Daddi et al.\ (2004) employs the $(B-z)$--$(z-K)$ color-color plane to separate $z\sim2$ galaxies from low-$z$ interlopers. We use a revised color-color criteria designed to select similar populations at $z\sim2$ based on $(g-z)$--$(z-K_s)$ and $(z-H)$--$(H-K_s)$ color-color planes to accommodate the bandpasses differences and the relatively shallow depth of the CFHT $g$-band data (Arcila-Osejo \& Sawicki 2013). In this paper, we describe the data set and simulation method, and present the measurements of angular correlation functions for $z\sim2$ galaxies sampled by a color-color selection method that closely resembles the $BzK$ technique. We defer the measurement of spatial correlation functions, as the deprojection of angular correlation functions requires robust estimates of the redshift distributions of the objects in question. We intend to incorporate photometric redshifts in a future paper. A standard concordance cosmology of $(\Omega_m,\Omega_\Lambda) = (0.3,0.7)$ and $H_0 = 70$ km s$^{-1}$ is assumed. All magnitude photometries are on the AB scale unless otherwise noted.

\section{Data and Catalog}\label{sec.2}

This study relies on visible/near-infrared imaging obtained at the Canada-France-Hawaii Telescope (CFHT). The visible $u^*g'r'i'z'$ data (which we abbreviate to $ugriz$ hereafter) come from the sixth release of the CFHT Legacy Survey (CFHTLS) Deep, covering four independent 1$\deg ^2$ fields (labeled D1, D2, D3, and D4) observed with MegaCam on CFHT. Most of these fields overlap with the regions of the sky that have been studied extensively in other surveys, including the COSMOS (field D2) and the Groth strip (D3); see, e.g., Gwyn (2011) and references therein. The image stacks with the 25\% best-seeing are used. The near-infrared $J H K_s$ imaging is from the T0002 release of the WIRcam Deep Survey (WIRDS; Bielby et al.\ 2012). The target fields are subsets of the CFHTLS fields, taken with the CFHT/WIRcam with the exception of the $J$ band image in D2, for which UKIRT/WFCAM data were used. The near-infrared camera has a smaller field of view than MegaCam, and the image for each field generally consists of mosaics. All images were processed by Terapix, astrometrically registered and resampled to a common pixel scale of 0.186$\arcsec$. The internal astrometric accuracy is $\sim$0.05\arcsec\ between the optical images and $\sim$0.1\arcsec\ with respect to the infrared.

\begin{figure}
\begin{center}
   \includegraphics[height=0.350\textheight]{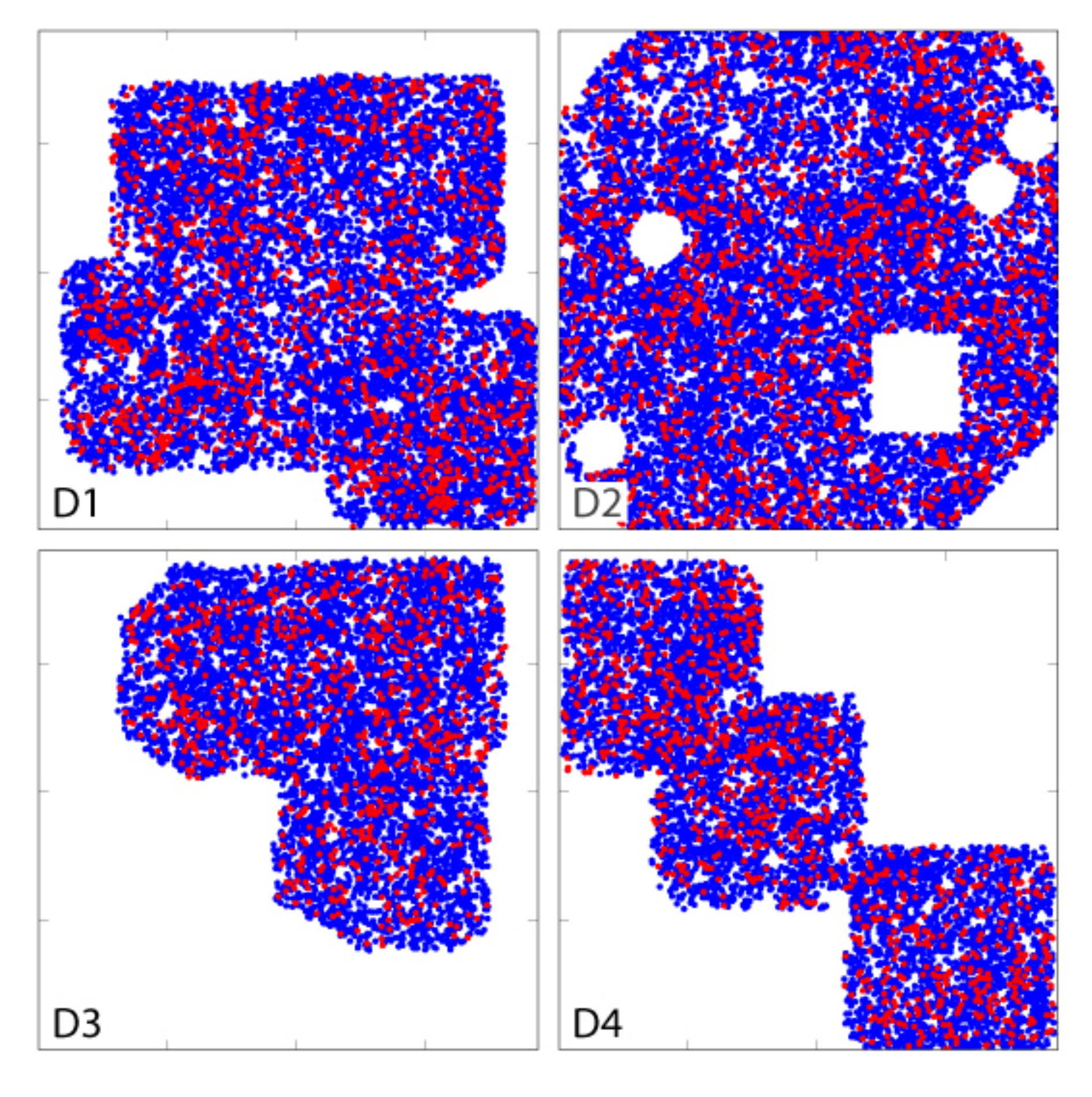}
\end{center}
 \caption[]{The survey areas showing the distributions of star-forming (blue) and passive (red) $gzHK_s$ galaxies. Each image extends over 19354 pixels on the side, which is about 1$\deg$ (i.e., 0.186$\arcsec$ per pixel). The regions with very poor S/N as well as those close to very bright stars and their diffraction patterns are masked and excluded from our analysis. The peculiar survey boundaries in D1, D3, and D4 are due to mosaicing patterns of the near-infrared pointings.}
\label{fig.1}
\end{figure}

Since the source catalogs are constructed from $K_s$ band images, the survey geometries are largely limited by the exposed regions in $K_s$, which are generally smaller (except for D2) and not square (Fig.~\ref{fig.1}). Image masks are generated to flag pixels occupied by stars, diffraction patters, and other blemishes. Unexposed pixels (based on weight images supplied by Terapix) are also masked. In addition, the regions in which suitable point sources for PSF matching (Appendix~\ref{sec:appendixA}) cannot be found are masked.

For the most part, the data products offered by Terapix are used throughout. To ensure that our fixed-aperture photometry samples a similar physical region for an object in all passbands, however, the point-spread functions (PSFs) are matched across all the passbands within each field by smoothing (Appendix~\ref{sec:appendixA}). The PSF-matched images are used for color photometry, yet the source detection is done on unsmoothed images for maximal detectability of faint sources.  Object detection, photometry, and the selection of $z\sim2$ galaxies is described in detail in Arcila-Osejo \& Sawicki (2013); below we summarize the main points.

\subsection{Photometric Measurements}\label{sec.2.1}

SExtractor (v2.8.6; Bertin \& Arnouts 1996) is used to generate object catalogs. For each field, source detection is carried out on the $K_s$-band image and then a set of fixed circular aperture photometry is done in the dual mode on sources identified in $K_s$. A source is considered detected when it extends over the minimum of 5 pixels at the 1.2$\sigma$ above the sky level. In addition to total magnitude estimates (via MAG\_AUTO), the color photometry is done through a 10-pixel diameter circular aperture, which corresponds to 1.86$\arcsec$ projected on the sky. Objects with SExtractor internal flags of $\le$ 4 are considered problem-free; objects flagged otherwise are treated as unobserved.

\subsubsection{The \gzHK\ Color Selection}\label{sec.2.1.1}

The objects of our interest are $z \sim 2$ galaxies which are color selected by a method devised to replicate the \bzk\ technique with the CFHTLS+WIRDS data (Arcila-Osejo \& Sawicki 2013), which we summarize here. Such a replication is done by tracing the population synthesis model tracks in the $BzK_s$ and $gzK_s$ color-color planes and redefining the demarcation lines in $gzHK_s$ so that the cuts select out similar galaxy populations as in the $BzK_s$. The distinction between star-forming and passive galaxies is made in the region where $(z-K_s) > 2.55$, where all objects are considered to be at $z \ga 1.4$. Due to the limited depth in the $g$ photometry (our substitute for $B$), however, we use supplemental information in near-infrared colors to better constrain their star formation history. 

For galaxies with $(z-K_s) \leq 2.55$, the ones that satisfy 
\[
(z-K_s)-1.27(g-z) \geq -0.022
\]
are classified as SF-\gzhk\ galaxies.  For objects with $(z-K_s) > 2.55$, SF-\gzhk\ galaxies satisfy the condition
\begin{equation*}
  (z - H) \leq 2.4 (H - K_s) + 1
\end{equation*}
and the rest are considered to be passive (PE-\gzhk). 

We stress that the \gzHKs\ technique is tuned to select the same galaxy populations as those selected by the classic Daddi et al.\ (2004) \bzk\ method. This work is thus fairly similar to most other Ò\BzKÓ studies that select high-redshift galaxies after transforming from their filter systems to the classic \BzK\ set. Consequently, the results of our \gzHKs\ selection are directly comparable to those from \BzK\ studies by design.

\subsubsection{Star-Galaxy Separation}\label{sec.2.1.2}

 In the $gzK_s$ plane, stars and galaxies are very clearly separated; this is in fact one of the advantages of the \bzk\ technique.  The objects which satisfy the criterion
\[ 
(z-K_s) - 0.45(g-z) \leq -0.57
\]
are considered stars. The rest are all considered galaxies. The term Òall galaxiesÓ is used in this paper to refer to the objects which do not satisfy the above criterion, which include SF-\gzHKs\ and PE-\gzHKs\ galaxies as well as lower-$z$ galaxies.

\section{RANDOM OBJECT SIMULATION}\label{sec.3}

A critical component of measuring angular correlation function is to compare the distribution of observed objects to that of the objects whose locations on the sky are uniformly randomized. Since this cannot be done with the real universe, one runs a set of simulations to create realizations of the universe from some model, with all underlying ÒparametersÓ tuned to approximate reality. For the purpose of clustering measurements, the object coordinates are uniformly randomized to see if the distribution of observed objects over the same survey area is more clustered compared to that of simulated universe with all objects uniformly distributed. A standard way to achieve this is to draw simulated objects from some priors that are based on theoretical models and/or observations. While the model dependence of this approach makes tractable the analysis and interpretation of results, it has a limitation in that it is a set of those underlying physical properties of the universe that one often seeks to constrain.

In our simulations we take a very empirical approach in which simulated objects 
are constructed to have the same properties as those in the catalog of observed objects, but with coordinate positions that are uniformly randomized. The method is still model-dependent in that the simulated objects are parametrically modelled (\S~\ref{sec.3.2}) but is adequate in that roughly similar proportions of galaxy types should be simulated to the extent that they are observationally distinguishable. For its simplicity this approach does not account at all for incompleteness in the observed catalog, in a sense that a luminosity function calls for\footnote{In our LF analysis (Arcila-Osejo \& Sawicki 2013) we gauge and correct for incompleteness by using the same simulations described here but selecting simulated objects that match \BzK\ galaxy properties in deep HST studies.}. The issue is mitigated by the fact that there is no simple way for carrying out incompleteness-corrected clustering measurements, since data-data pair separations can only be computed from what is observed. So long as the analysis remains on reasonably complete catalogs, the lack of such correction should not render our results useless. The incompleteness in the observed object catalog and appropriate corrections necessary for simulations are discussed in \S~\ref{sec.3.3}.

To account for systematics inherent in real observations, the random object catalog must be subject to the same detection procedure as the one used in the real observations. Briefly, the procedure is as follows: (1) detect sources in an observed image to construct the data catalog; (2) draw objects randomly from the data catalog and implant them onto a background image to construct a simulated image; (3) run the same detection process on the simulated image; and finally (4) match the simulated and recovered objects. The (master) random object catalog is generated from a large number of realizations (400), each constructed from the procedure just outlined. On the computations of pair counts for angular clustering, the random objects are bootstrap-resampled from this catalog. The details will be discussed in the following sections.

\subsection{Definition of Photometric Recovery}\label{sec.3.1}

An essential step in photometric recovery of simulated objects (\S~\ref{sec.3.2}) is to subject them to the same observational/instrumental systematics, to the same detection process as is used in generating the catalog of observed objects. Such a recovery procedure naturally takes care of common systematics due to survey geometry and varying background noise characteristics over the survey field. When implanted onto an observed image, there may be an additional complication in the recovery process due to overlapping/crowding with existing sources. To better isolate the effect of the background noise, a set of simulated blank sky images were created from observed images. Object mask images generated by SExtractor were used to identify source regions, which were then filled with artificial pixel values with variances scaled to local backgrounds. Such images are useful for inspecting recovery rates without the complications caused by overlapping with objects in observed images.
\begin{figure}
\begin{center}
  \includegraphics[height=0.350\textheight]{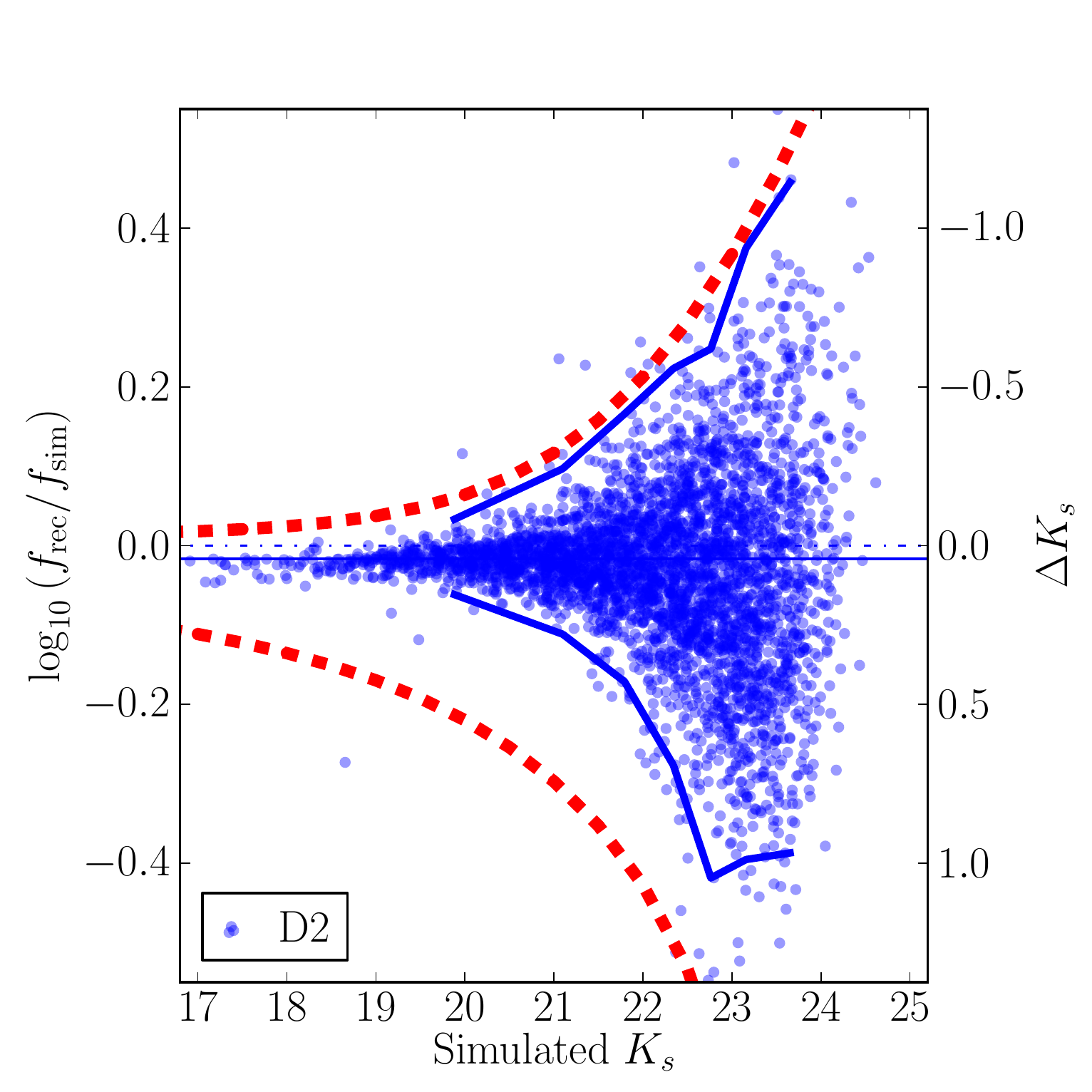}
\end{center}
 \caption[]{The logarithmic ratios of recovered ($f_{rec}$) and simulated ($f_{sim}$) fluxes as a function of simulated $K_s$ magnitudes in D2. The objects are implanted onto a simulated background image. The solid blue curves indicate 97.5\% bounds within sliding bins. The red dotted lines indicate the bounds used for photometric match (\S~\ref{sec.3.1}). The solid line just below $\log_{10} (f_{rec}/f_{sim})=0$ roughly indicates the median of the measurements (for bright sources), implying there is some loss in recovered flux via SExtractor measurements. Repeating this exercise with the observed image as the background will produce a number of recovered fluxes well above the upper bound for photometric matching, indicating overlapping between simulated objects can significantly affect recovery rate. To reduce the file size, the number of objects shown are thinned out to one quarter of the original..}
\label{fig.2}
\end{figure}

To ensure that recovered objects are indeed the ones implanted, we rely on the discrepancies between the simulated and recovered positions as well as Ks magnitudes. First, matching between simulated and recovered catalogs by positions is done through $k$-d tree nearest-neighbour lookup with the maximum distance of 10 pixels, large enough for drifting centroids of most objects at the detection threshold. The ID number of the closest object detected is associated with the simulated object. Second, the simulated and recovered fluxes are compared, and if they are within the magnitude-dependent bounds (Fig.~\ref{fig.2})\footnote{When presenting figures, we opt to do so with the results in the D2 field. The choice is made to facilitate comparisons with literature where relevant, mainly guided by the fact that the field overlaps with the extensively-studied COSMOS survey (e.g., McCracken et al.\ 2010). Note, however, that the D2 field is the shallowest (in $K_s$) of all, and the PSFs were matched to the worst seeing in $J$ for color photometry; consequently, in the other three fields various effects will set in at $\sim$1 mag fainter than in D2.}, the object is tagged as photometrically-matched. The photometric cut is fairly conservative and does not aggressively eliminate misidentifications, especially at faint magnitudes. It is done nonetheless to reduce gross misidentifications due to overlapping and errant measurements. Fig.~\ref{fig.2} is also useful to assess the number of recovered objects ÒleakingÓ in or out to nearby magnitude bins due to photometric scatter, especially important near the magnitude limit. At and below $K_s \approx 23$, a significant fraction of objects are recovered at more than $\Delta K_s = \pm 0.5$ away from simulated magnitudes, which introduces Eddington bias. The significance of this effect varies depending on binning and photometric errors (Teerikorpi 2004), and should be taken into account in correcting for statistically missing faint populations as in number counting. Some studies of faint galaxy number counts --- including ours presented in Arcila-Osejo \& Sawicki 2013 --- which do not take this effect into account may slightly undercorrect for fainter populations. Also noticeable in the figure is the flux loss of about a few percent between simulated and recovered objects, which we correct for subsequent simulations.

\begin{figure}
\begin{center}
  \includegraphics[height=0.350\textheight]{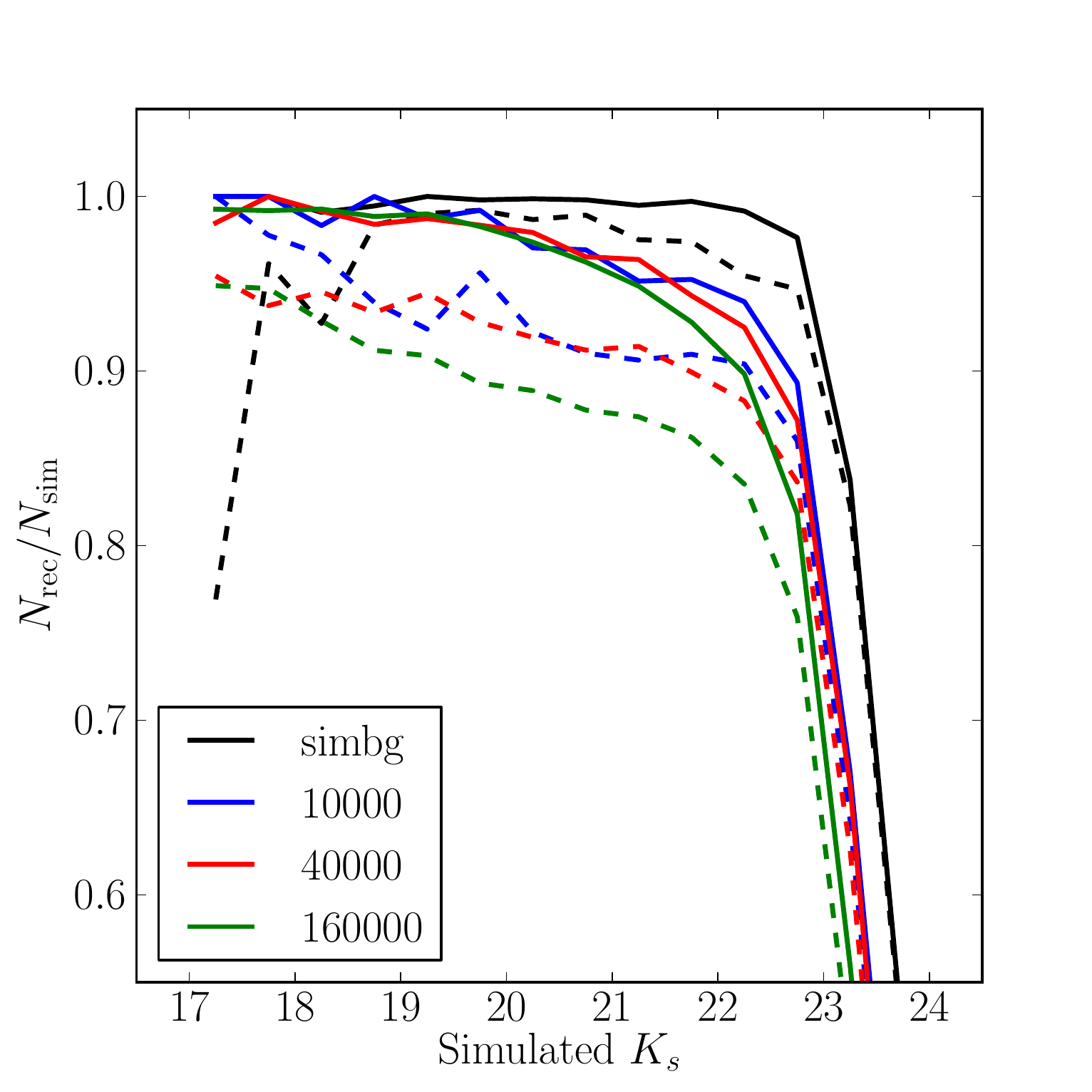}
\end{center}
 \caption[]{
 The object recovery fraction as a function of simulated $K_s$ magnitude in D2. Solid lines are completeness functions when object recoveries are based only on their coordinates, whereas dashed lines are for recovery based both on coordinates and photometry; see \S~\ref{sec.3.1}. In the black lines, we see recovery function when simulated objects are implanted onto a simulated background image. Different numbers of simulated objects are implanted onto the D2 field image: 10,000 (blue), 40,000 (red), and 160,000 (green) to see where overlapping between simulated objects starts affecting object recovery.
 }
\label{fig.3}
\end{figure}

In ÒrealÓ simulation runs, the objects are implanted onto the observed images. This is done to reflect the survey geometries and noise properties, but also to take into account the overlapping of the objects of interest at $\z\sim 2$ with foreground/background sources. Crowding the field too much with simulated objects, however, will introduce unwanted systematics due to overlapping among simulated objects themselves\footnote{Ideally, a better ÒbackgroundÓ image can be constructed by removing only the objects of interest from the field. Implanting simulated objects of interest into such an image would better take into account the systematics caused by object collisions between foreground/background sources.}. The optimal number of simulated objects to be added to an image at once is found by inspecting the drop in the recovery rate as a function of the number of simulated objects (Fig.~\ref{fig.3}). Compared to the recovery rate for the case when simulated objects are added onto a simulated background image, all recovery rates of the objects on observed images suffer from the loss of objects due to overlapping. In addition, the recovery rate generally drops ($\sim$10\%) when an additional check on photometric match between simulated and recovered magnitude is carried out in addition to position match. Overlapping with existing objects (and other simulated objects) causes both/either the centroids to move and/or the fluxes to be bumped up for the simulated objects upon recovery. Hence the reductions in recovery rates as seen in Fig.~\ref{fig.3} are expected. Nonetheless, little change in recovery function is observed up to 40,000 simulated objects per simulation run per image. To be conservative, no more than 20,000 objects are implanted onto a science image at a time in our simulations.

\subsection{Simulated Galaxy Model}\label{sec.3.2}

\begin{figure}
\begin{center}
  \includegraphics[height=0.350\textheight]{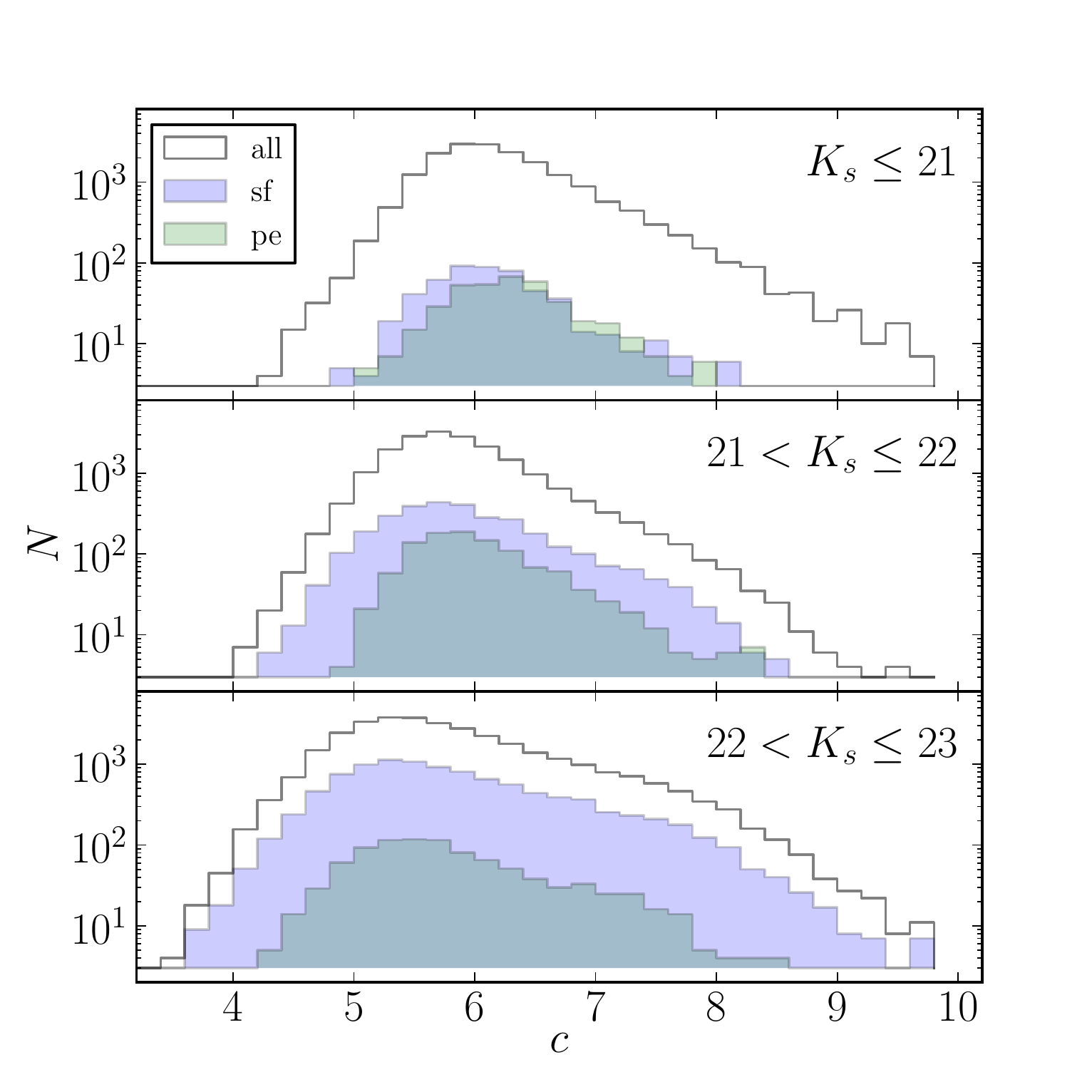}
\end{center}
 \caption[]{
The distributions of $K_s$-band concentration indices by $K_s$ magnitudes for SF-\gzHK\ (blue histogram), PE-\gzHK\ (green), and all galaxies (white). This figure only shows the observed objects in the D2 field, but a similar trends are observed in the other fields. }
\label{fig.4}
\end{figure}

\begin{figure}
\begin{center}
  \includegraphics[height=0.350\textheight]{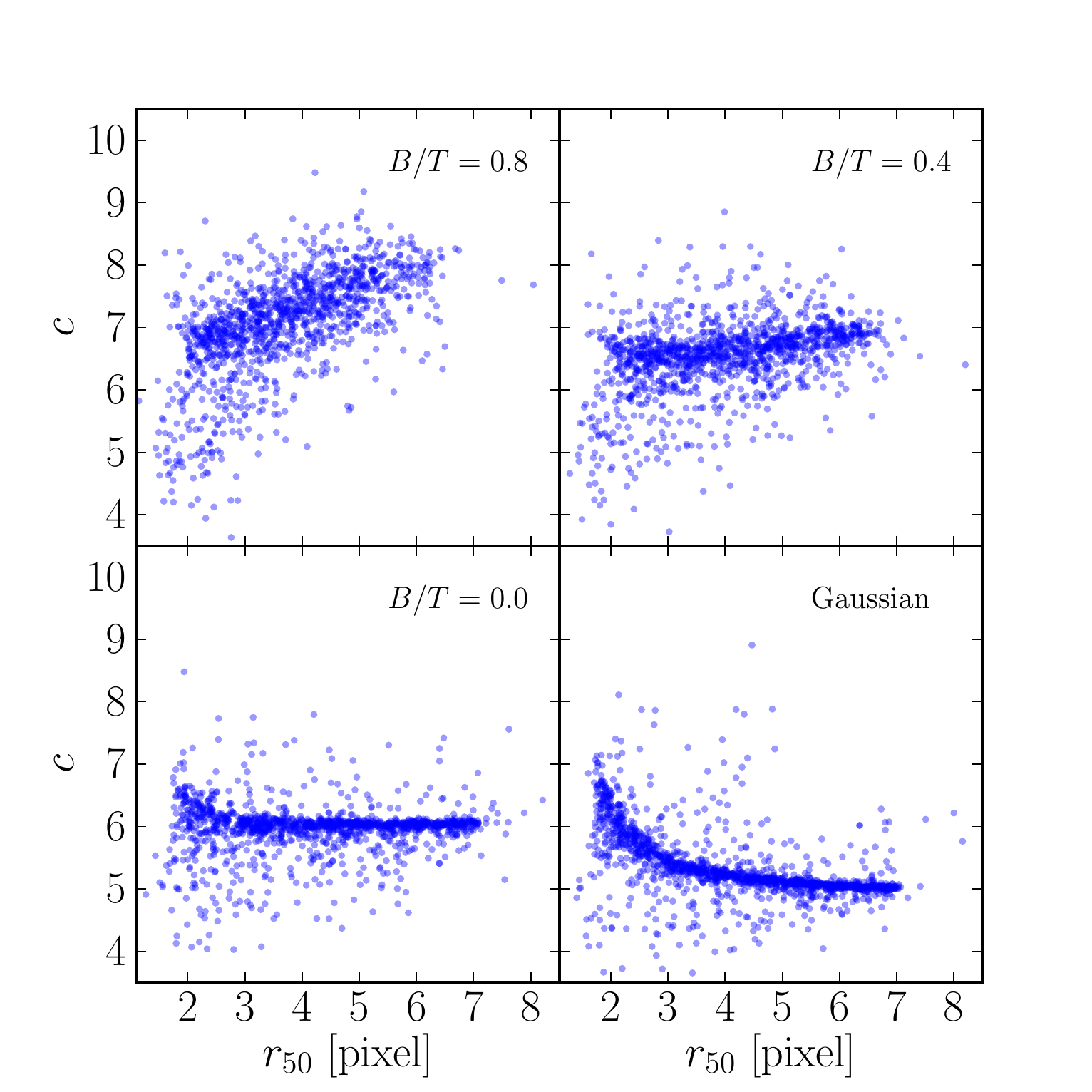}
\end{center}
 \caption[]{
The distributions of measured $K_s$-band concentration indices $c$ and half-light radii $r_{50}$ for input models with different bulge-to-total light fractions (B/T) in D2. The parametric model used for simulated galaxies is described in \S~\ref{sec.3.2}. The distribution for Gaussian model for point sources (bottom right) is also shown for reference. For clarity, the number of objects shown is thinned out to one tenth of the original. }
\label{fig.5}
\end{figure}

We model a galaxy as a combination of disk and de Vaucouleur (i.e., Sersic with n = 4) spheroid, a parameteric model similar to GIM2D (Simard et al.\ 2002) with the following constraints. Same position angles are assumed for disk and spheroid. The effective radii (i.e., half-light radii) of disk and spheroid are assumed to be similar. Bulge ellipticity and the disk inclination are coupled. In practice, the objects of interest are barely resolved, so these technicalities do not affect our science.
While the image quality does not allow us to carry out detailed morphological analysis for $z\sim 2$ objects, we attempt to incorporate some morphological information into our simulation via the concentration index $c$, as defined by Kent (1985):
\[
c = 5 \log{\left(r_{80} / r_{20}\right)}
\]
where $r_{80}$ and $r_{20}$ are the radii enclosing 80\% and 20\% of total flux, respectively. Fig.~\ref{fig.4} shows the distributions of concentration indices for galaxies grouped by their $K_s$ magnitudes and types. For $K_s$-bright objects, PE-\gzHKs\ galaxies are slightly more concentrated than SF-\gzHKs\ galaxies in general, but the distinction weakens as they become fainter and the morphological information gets lost. As shown in Fig.~\ref{fig.5}, however, input models do affect the measured concentration indices; bulge-dominated objects tend to get recovered as highly concentrated objects. In our parametrization of simulated object, the bulge-to-total light fraction (B/T) is the primary parameter controlling the light concentration. Using Fig.~\ref{fig.5}, we empirically map the measured concentration index and half-light radius onto B/T.

\begin{figure}
\begin{center}
  \includegraphics[height=0.350\textheight]{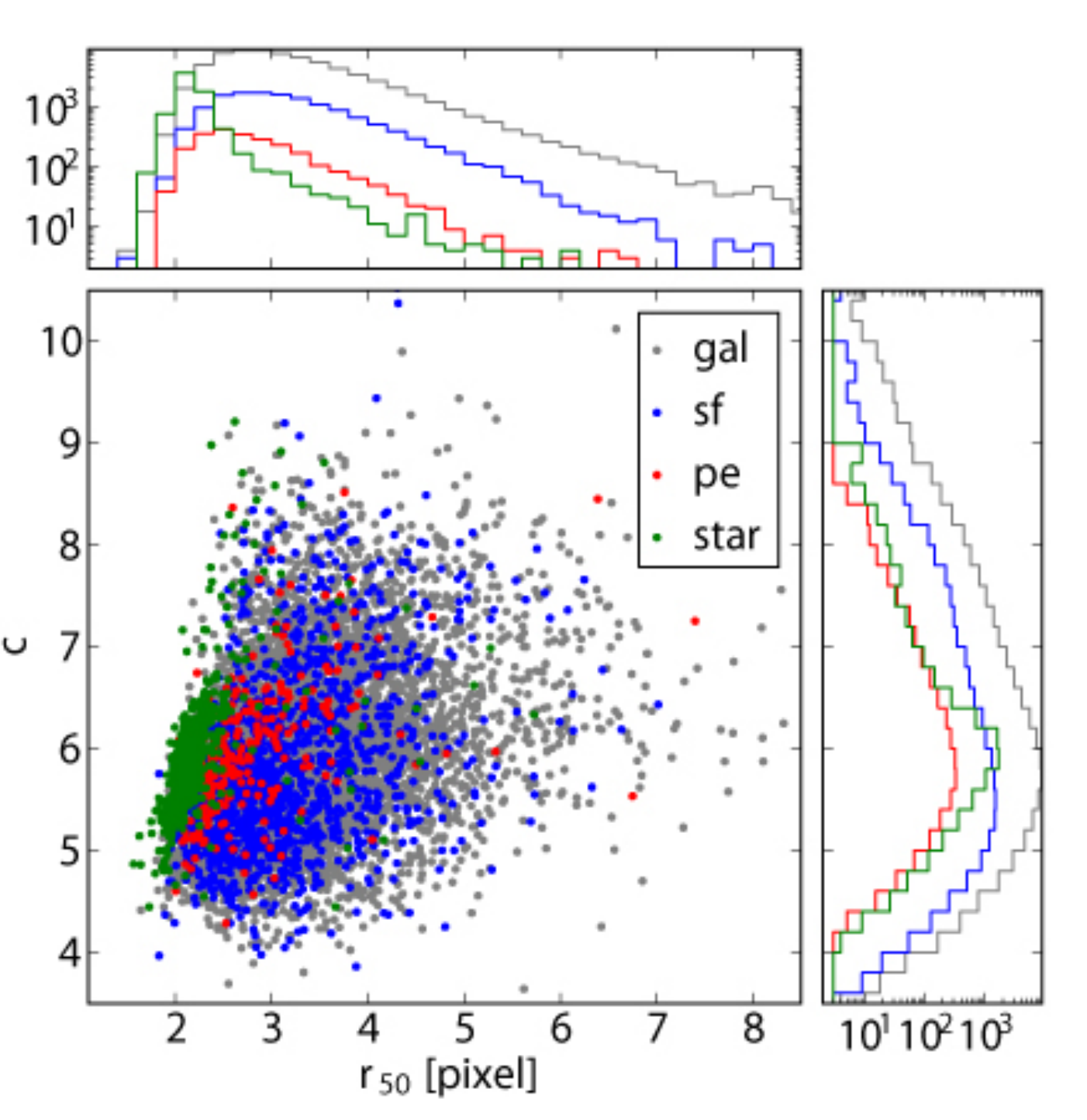}
\end{center}
 \caption[]{
The distributions of $K_s$-band concentration indices $c$ and half-light radii $r_{50}$ for $K_s \leq 23$ objects in D2. In the scatter plot and histograms, 
SF-\gzHK\ galaxies are coloured blue, PE-\gzHK\ red, all other galaxies are gray, while stars are green. For clarity, the number of objects is thinned out to one tenth of the original. }
\label{fig.6}
\end{figure}

\begin{figure}
\begin{center}
  \includegraphics[height=0.350\textheight]{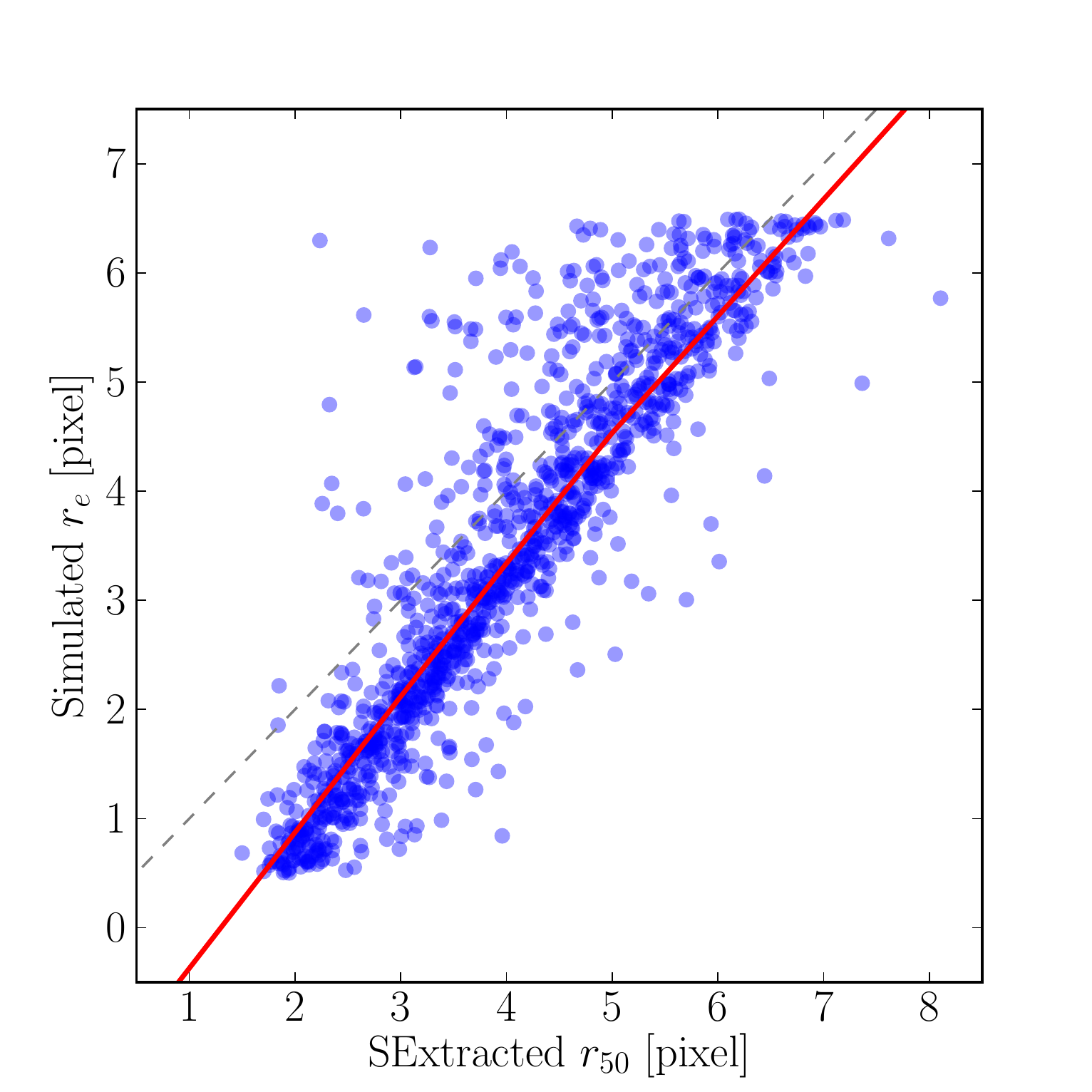}
\end{center}
 \caption[]{
The input model effective radius $r_e$ as a function of the measured half-light radius $r_{50}$ in D2. The dotted line denotes unity where $r_e = r_{50}$. The red solid curve is a fiducial fit to map the measured half-light radius to input model effective radius in our simulations. The curve is chosen mostly to go through the dense parts of distribution, but manual adjustments are made to better match the distribution of $r_{50}$ toward the smaller end in real simulation runs (Fig.~\ref{fig.8}), i.e., it has not been determined via a rigorous regression analysis. For clarity, the number of objects plotted is thinned out to one tenth of the original. }
\label{fig.7}
\end{figure}

While the distributions of concentration indices are not distinguishable for all but the brightest \gzHKs\ galaxies, their size distributions are clearly different (Fig.~\ref{fig.6}). The PE-\gzHKs\ galaxies are generally smaller than SF-\gzHKs\ galaxies in terms of half-light radius ($r_{50}$). Since the objects are broadened due to seeing and instrumental systematics, the measured $r_{50}$ are mapped to the model effective radius $r_e$ empirically using Fig.~\ref{fig.7}, recovering simulated objects with varying model parameters. This is certainly simplistic and does not take into account such obvious complications as inclination effects, but we find it does not significantly affect completeness (\S~\ref{sec.3.3}).

\subsection {Completeness Correction}\label{sec.3.3}

The simulated objects are randomly drawn directly from the observed object catalog in an attempt to reproduce the same mix of objects. Corrections still need to be applied to the populations which suffer significant incompleteness; otherwise, the random object catalog will suffer twice the incompleteness --- once when an observed object catalog is constructed and yet another when the objects are drawn from the object catalog, implanted onto a background image, and (un)recovered in simulation.

\begin{figure}
\begin{center}
  \includegraphics[height=0.350\textheight]{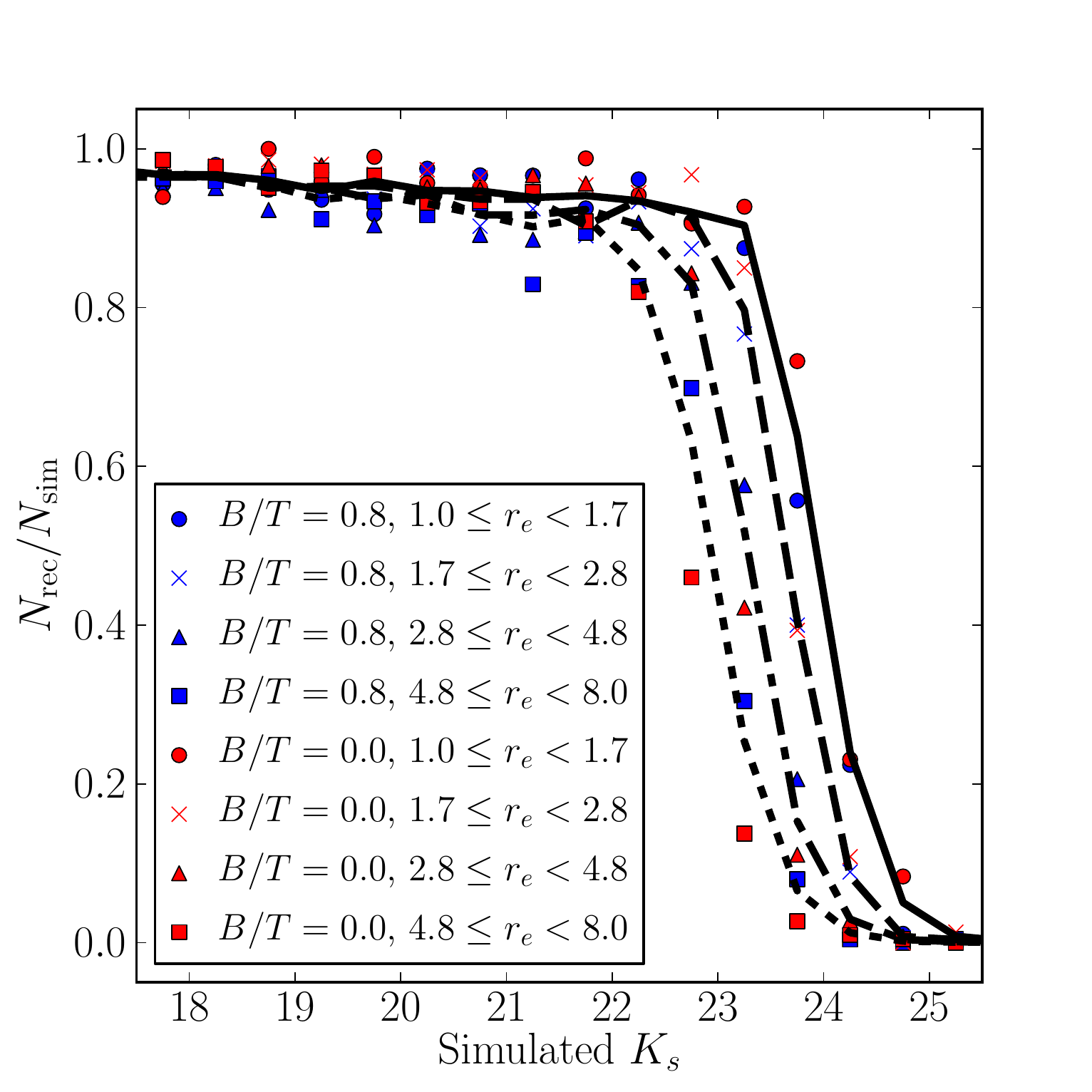}
\end{center}
 \caption[]{
The object recovery fractions as functions of simulated $K_s$ magnitudes, separately computed for different model effective radii in D2. The black lines indicate recovery functions for $1 \leq  r_e < 1.7$ (solid), $1.7 \leq r_e < 2.8$ (dashed), $2.8 \leq r_e < 4.8$ (dash-dotted), and $4.8 \leq re < 8.0$ (dotted) with $r_e$ in units of pixels. The points indicate recovery fractions computed separately for model bulge fractions B/T = 0.8 (blue) and 0 (red), and for $1 \leq r_e < 1.7$ (circle), $1.7 \leq r_e < 2.8$ (cross), $2.8 \leq r_e < 4.8$ (triangle), and $4.8 \leq r_e < 8.0$ (square). }
\label{fig.8}
\end{figure}

The primary factors that affect the recovery rate of an object are the brightness and the light profile of that object. The completeness functions were inspected for simulation model parameters that affect these aspects of photometry. The ellipticity $e$ of simulated object does not contribute significantly to incompleteness; the objects with ellipticity $e \ga 0.8$ cannot be recovered by SExtractor, and the completeness at $e \la 0.8$ does not vary (plots not shown). Aside from the total magnitude, it is the effective radius that contributes most significantly to the detectability of objects; see Fig.~\ref{fig.8}. The same figure also indicates that the model B/T does have a significant effect on the recovery rate toward faint magnitudes. However, morphological information gets lost at faint magnitudes (Fig.~\ref{fig.4}) and therefore we may not be able to correct for the effect.

\begin{figure}
\begin{center}
  \includegraphics[height=0.350\textheight]{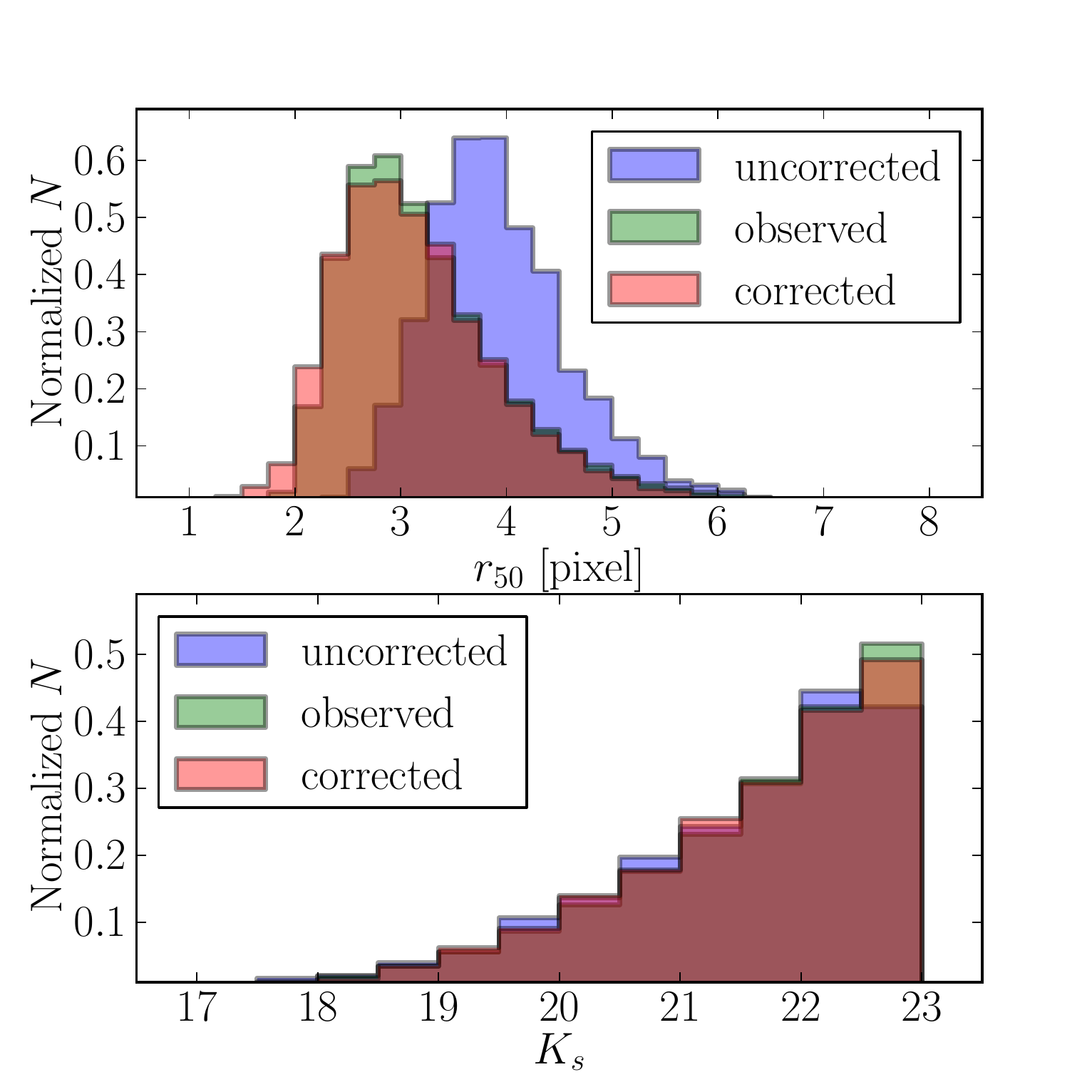}
\end{center}
 \caption[]{
 The normalized histograms for measured half-light radii $r_{50}$ (top) and $K_s$ magnitudes for galaxies in D2. The distributions are shown for the observed catalog (green), the random object catalog uncorrected for incompleteness (blue), and the random catalog corrected for incompleteness (red) as described in \S~\ref{sec.3.3}.
 }
\label{fig.9}
\end{figure}

Finally, we compare the random object catalog to the observed catalog to see how well the former reproduces the latter. As seen in Fig.~\ref{fig.9}, the observed and completeness-corrected catalogs compare well in terms of $K_s$ magnitude and $r_{50}$ distributions. The figure shows that without corrections we would be missing a large number of objects with small half-light radii $r_{50}$, the class to which a large number of \gzHKs\ galaxies belong (Fig.~\ref{fig.6}).

\section{CLUSTERING PROPERTIES OF $z \sim 2$ GALAXIES}\label{sec.4}

\subsection {Two-Point Angular Correlation Function}\label{sec.4.1}

The two-point angular correlation function $\omega (\theta)$ is computed using the Landy \& Szalay estimator (Landy \& Szalay 1993):
\begin{equation}
  \omegals\tp =
  \frac{r (r - 1)}{n (n - 1)} \frac{\DD\tp}{\RR\tp}
  - \frac{r - 1}{n} \frac{\DR\tp}{\RR\tp}
  + 1 \ ,
    \label{eqn:1}
\end{equation}
where $DD(\theta)$, $DR(\theta)$, and $RR(\theta)$ are the numbers of data-data, data-random, and random-random pairs with angular separations between $[ \theta-\delta\theta/2, \theta+ \delta\theta/2)$. The total numbers of objects in the data ($n$) and random ($r$) catalogs are used to normalize the pair counts in the above expression. The estimator has become the de facto standard for computing angular correlation functions in galaxy surveys.

For a known angular correlation function $\omega(\theta)$, the number of pairs with separations in  $[ \theta-\delta\theta/2, \theta+ \delta\theta/2)$ is given by
\[
  n
  \left(\frac{\delta\Omega_1}{\Omega}\frac{\delta\Omega_2}{\Omega}\right)
  \left[1 + \omegat\right] \ ,
\]
Doubly integrating this over the solid angles $\Omega_1$  and $\Omega_2$ for
the entire survey area should recover n, as the total number of
unique data-data pairs is a fixed quantity; a strong clustering
signal at small angular separations must be
balanced by a weak clustering signal at large angular separations. However, the normalization of $\omega (\theta)$
depends on the survey geometry as well as how well-sampled
the clustering signal is. For example, if the survey area is so
small that a measurement only captures a clustering signal
of a high variance over the limited region, $\omega(\theta)$ measurement
tends to underestimate the true angular correlation. Suppose $\omega_{est}(\theta)$
is computed from some estimator (e.g., $\omega_{LS}$), and $\omega(\theta)$ is the ``true" 
angular correlation function for the sample, such that 
\begin{equation}
  1 + \omega_{\rm est}\tp = f \left(1 + \omega\tp\right) \ ,
\label{eqn:2}
\end{equation}
where $f$ is a factor that puts $\omega(\theta)$ on a similar scale with $\omega_{est}(\theta)$. The following constraint
\[
  n = \iint n \left(\frac{\delta\Omega_1}{\Omega}\frac{\delta\Omega_2}{\Omega}\right)
  f \left[1 + \omega\tp\right] \ ,
\]
leads to
\[
  f = \left[ 1 + \iint \omega\tp \frac{\delta\Omega_1}{\Omega}\frac{\delta\Omega_2}{\Omega} \right]^{-1} = \frac{1}{1 + \ic} \ ,
\]
where $\ic$ is an integral constraint:
\begin{equation}
  \ic \equiv \frac{1}{\Omega^2} \iint \omega\tp d\Omega_1 d\Omega_2 \ .
\label{eqn:3}
\end{equation}
For $\ic \ll 1$, \eq{eqn:2} suggests
\[
  \omega\tp \approx \omega_{\rm est}\tp + \ic \ ,
\]
which leads to the earlier point on how $\omega_{set}$ tends to underestimate the true angular correlation function\footnote{
This discussion followed the presentation by Wall \& Jenkins (2003). For another perspective, see Adelberger et al.\ (2005), for example.
}.  Since the true form of $\omega(\theta)$ is not known, a power law of the form 
\begin{equation}
\omega(\theta) = A_\omega \theta^{1-\gamma}
\label{eqn:4}
\end{equation}
is often assumed. Using this expression in \eq{eqn:3} leads to an
estimate of integral constraint
\begin{equation}
  C \equiv \ic / \Aw = \frac{\sum \RR\tp \theta^{1 - \gamma}}{\sum \RR\tp}
    \label{eqn:5}
\end{equation}
\citep{roch99} such that
\begin{equation}
  \omegat = \Aw \left( \theta^{1-\gamma} - C \right) \ .
  \label{eqn:6}
\end{equation}
The correlation functions computed via \eq{eqn:1} from the data
and random catalogs are often fitted by \eq{eqn:6}.  It should be
noted that the term integral constraint appears to be used for both
\ic\ and $C$ interchangeably in the literature.  Henceforth we assume the
form of $C$ when integral constraint is discussed in this paper.

Uncertainties in pair counts over different angular separations are obtained by bootstrap resampling of objects in the data (i.e., observed) and random catalogs (\S~\ref{sec.3}). At each bootstrap realization, a new data catalog is generated by randomly resampling (with replacement) the same number of objects from the original data catalog. The random catalog is constructed similarly. The angular separation histograms $DD$, $RR$, and $DR$ are computed between all unique pairs and are binned up in the range $-4.5 \leq \log_{10} \theta[deg] \leq +1.0$, at a logarithmic interval of $\delta\theta = 0.2$ dex. The correlation function is computed via \eq{eqn:1}. The bootstrap simulation is repeated about 100 times. Since the value of $\omega(\theta)$ within each angular separation bin is found to be normally distributed for each field, the mean and standard deviation are computed in the standard manner to obtain the best estimate and uncertainty for each bin. The average of $RR$ at each angular separation bin is also recorded for the purpose of estimating the integral constraint by means of \eq{eqn:5}. By nature of clustering measurement, $\omega(\theta)$ in different bins are in fact correlated, but we do not take this into account in our uncertainty estimates. For an approach to estimate full covariance, see Wake et al.\ (2011), for example.

\subsection{Fit Parameters and their Systematic Biasing}\label{sec.4.2}

The parametric function of the form in \eq{eqn:6} is fitted to the angular correlation functions as outlined in \S~\ref{sec.4.1}, each bin weighted by its inverse variance. The clustering amplitude $A_\omega$, slope $\gamma$, and integral constraint $C$ are estimated via the standard Markov-chain Monte Carlo (MCMC) sampling technique as implemented by PyMC (Patil et al.\ 2010). Due to the form of \eq{eqn:5} and the fixed set of random-random pairs $RR$ (\S~\ref{sec.4.1}), there is a one-to-one relation between $C$ and $\gamma$ when these two parameters are both free.

\begin{figure}
\begin{center}
  \includegraphics[height=0.320\textheight]{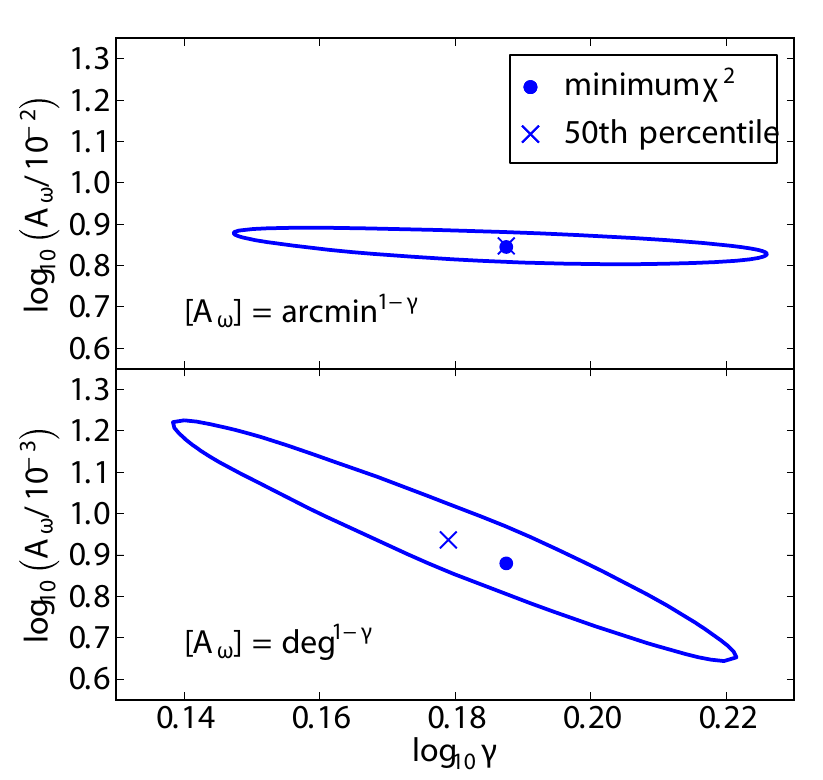}
\end{center}
 \caption[]{
The 68\% posterior probability contours of model parameters \Aw\ and \g\ of the correlation function for SF-\gzHKs\ galaxies in D1, as estimated by MCMC sampling technique (\S~\ref{sec.4.2}). In \eq{eqn:6}, the angular separations are in units of arcminutes (upper) and degrees (lower). Uniform probability density priors are assumed for both parameters. Circular points indicate the parameters at the minimum \chisq. Crosses indicate the parameters at the 50th percentiles of MCMC posterior distributions. }
\label{fig.10}
\end{figure}

In using the fitting equation of the form in \eq{eqn:6}, it is often mentioned that the iterative approach required to estimate an integral constraint from \eq{eqn:5} leads to an unstable solution (e.g., Adelberger et al.\ 2005; Blanc et al.\ 2008). Most studies therefore opt to compute the integral constraint independently either by assuming a fiducial slope $\gamma$ (often $\gamma = 1.8$) or from a theoretical halo model and keeping it fixed for a sub-sample and given survey field. We argue that the lack of convergence via \eq{eqn:5} and \eq{eqn:6} may at least in part be caused by the particular numerical method that's employed. In Fig.~\ref{fig.10}, we show the posterior distributions of model parameters  $A_\omega$ and $\gamma$ for two similar yet technically different cases of MCMC sampling on the same data. The amplitude $A_\omega$ in \eq{eqn:6} is defined at where the term $(\theta^{1-\gamma}-C)$ is unity. In the case where $\theta$ is in units of arcminutes, only a very weak correlation is observed between $A_\omega$ and $\gamma$. When $\theta$ is in degrees, on the other hand, a strong correlation becomes apparent between the two model parameters; the best estimates for $A_\omega$ and $\gamma$ from the minimum-$\chi^2$ and at the 50th percentiles of posterior distributions also differ quite significantly. The correlation becomes even stronger with $\theta$ in radians (not shown). Effectively, we confirm the nonconverging tendency and find that estimating $A_\omega$ and $\gamma$ (and $C$ coupled to $\gamma$ via \eq{eqn:5}) simultaneously as free parameters may lead to systematically lower estimates for $\gamma$  when the clustering amplitude $A_\omega$ is defined in certain angular units. Yet, all these cases supposedly are mathematically similar, so the differences may be attributed to how a particular numerical method spans the parameter space to attain convergence. In other studies, it is often unspecified how $A_\omega$ is normalized; however, their clustering data must not be very well sampled over degree scales, so it is possible that they suffer from similar artificial numerical issues when $A_\omega$ is defined at those angular scales.

Since our clustering analyses are most relevant on arcminute scales, the angular separations in units of arcminutes are used in \eq{eqn:5} and \eq{eqn:6}; hence $A_\omega$ technically is in units of arcmin$^{1-\gamma}$ (so that $\omega$ remains dimensionless). The priors for $A_\omega$ and $\gamma$ are uniform probability densities with the range wide enough not to truncate the posterior distributions at extreme values. In all cases the posterior distributions are observed to be roughly lognormal, and their 50th percentiles mostly match the best estimated parameters at minimum \chisq.

In this paper, we generally quote the parameters at minimum \chisq\ as the best estimates. The 16th and 84th percentile bounds in the marginal distributions of parameters are quoted as the estimates for uncertainties in those parameters. The percentiles are chosen to roughly match commonly-cited 1$\sigma$ uncertainty in literature. We reiterate that the posterior distributions are roughly lognormal.

\subsection{Combined Angular Correlation Functions}\label{sec.4.3}

 The distribution of $\omega_{LS}$ in each angular separation bin is roughly normal (\S~\ref{sec.4.1}) and we combine the measurements from our four independent, widely-separated fields, to arrive at our best estimates of angular clustering as follows.  First, before combining $\omega$ from our four fields, we correct them for integral constraints (\eq{eqn:6}). Next, we compute the weighted mean $\bar\omega$ of the angular correlation functions, where inverse variance $w = 1/\sigma^2_\omega$ is used as the weight. The variance of the weighted mean is computed from
\[
\sigma^2_{\bar\omega}=
\frac{1}{\Sigma^n_{i=1} w_i}
\times
\frac{1}{n-1}
\sum_{i=1}^n  w_i (\omega_i - \bar\omega)^2, 
\]
where $n = 4$ for the four fields. We then have the combined-field angular correlation function $\bar\omega$ and the standard deviation $\sigma_{\bar\omega}$ for each angular separation bin. The clustering amplitudes and slopes in Table 1 are estimated from fitting the power law \eq{eqn:6} (but with $C = 0$) to the combined-field correlation function.

\begin{figure}
\begin{center}
  \includegraphics[height=0.340\textheight]{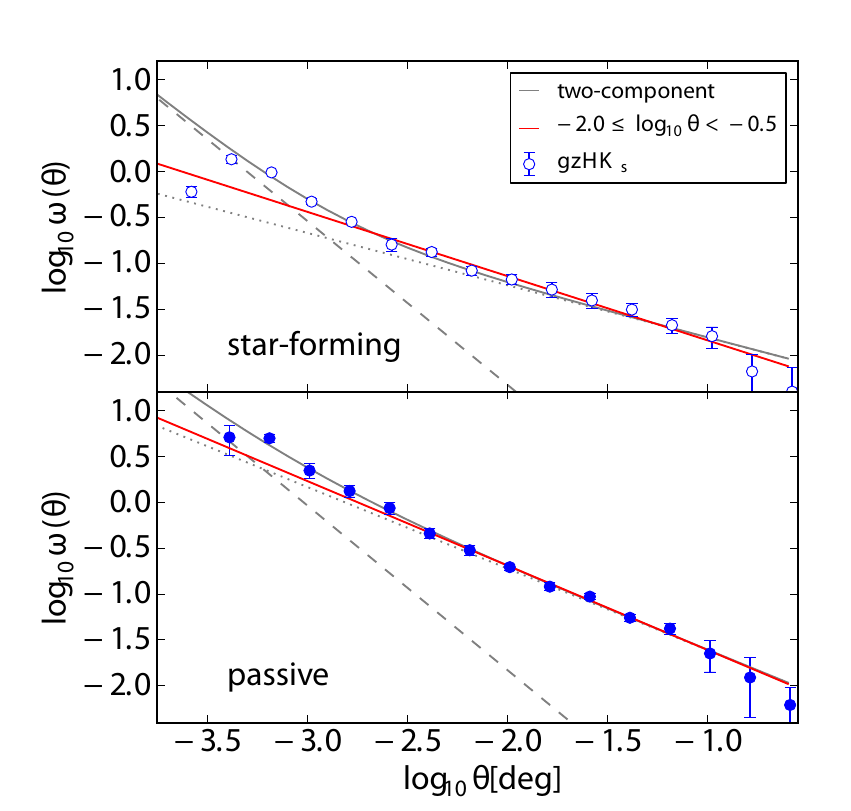}
\end{center}
 \caption[]{
The angular correlation functions as a function of angular separation for SF-\gzHKs\ (upper panel) and PE-\gzHKs\ (lower panel) galaxies. Each correlation function is fitted via \chisq\ minimization with a two-component power law of the form \eq{eqn:6}, ensuring one of the slopes \g\ is always larger than the other. The components with a flatter slope (dotted curve) and with a steeper slope (dashed curve) are linearly added, which gives an overall better fit (solid curve) to the observed angular correlation function. The fitting is done over the interval of $-3.5 < log_{10} \theta < -0.5$. The best fit curve from a single-component fitting over $-2.0 \leq log_{10} \theta < -0.5$ is also shown as reference (red curve). }
\label{fig.11}
\end{figure}

In recent years, it has been realized that angular correlation functions are better modeled by a two-component power law that includes a two-halo term from large-scale clustering of galaxies hosted in separate dark halos and a one-halo term from objects occupying the same halo. These components are apparent in sufficiently high-quality data (e.g., Zehavi et al.\ 2004; Wake et al.\ 2011). We also see evidence for the ÒbreakÓ caused by the two components in our data (Fig.~\ref{fig.11}). For both SF-\gzHKs\ and PE-\gzHKs\, the breaks appear around $log_{10} (\theta/1 deg) \approx -2.7$, which corresponds to $\sim$60~(physical)~kpc at $z=2$ in our adopted cosmology.  The projected angular separation at the break is well below the virial radius ($r_{200}$) of a $\sim10^{13}$\Msun\ halo, roughly the most massive at $z\sim2$, so the enhancement of clustering at the small angular scale is expected from one-halo contributions.

\begin{figure}
\begin{center}
  \includegraphics[height=0.320\textheight]{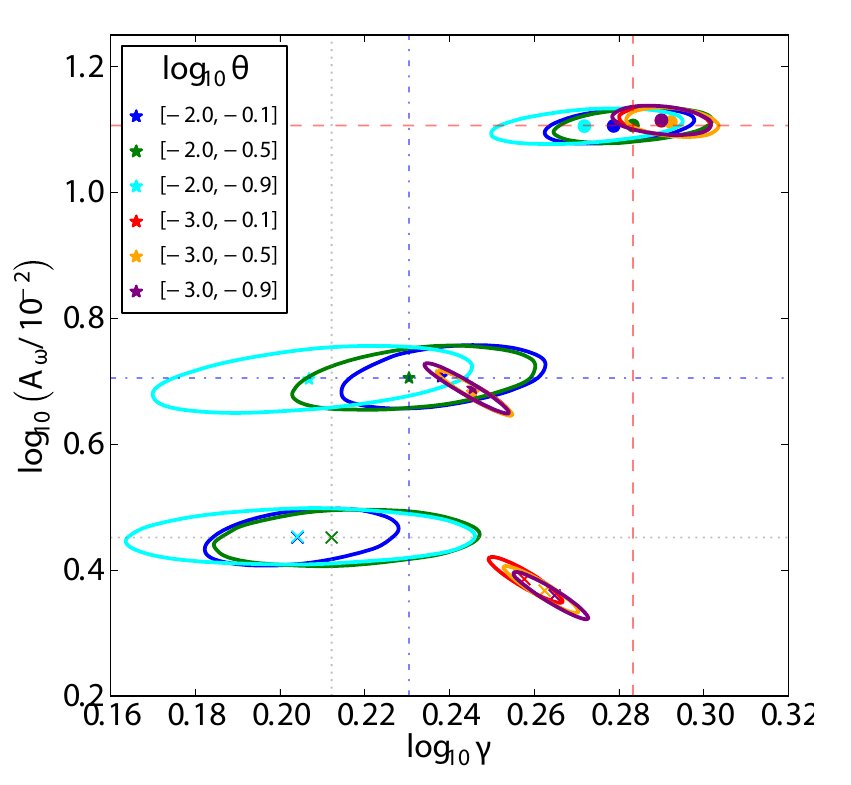}
\end{center}
 \caption[]{
The variation of the 68\% confidence intervals of the clustering amplitude \Aw\ and the power-law slope \g\ for PE-\gzHKs\ (upper contours), SF-\gzHKs\ (middle contours), and all galaxies (lower contours) when the fitting angular separation intervals are varied; see Table~\ref{tab:1}. The colors of contours indicate the fitting interval as indicated in the legend, except the gray contours which are from fitting without integral constraints. The best parameter estimates from the minimum-\chisq\ are indicated for PE-\gzHKs\ (points), SF-\gzHKs\ (stars), and all galaxies (crosses). The red dashed lines indicate the best estimates of \Aw\ and \g\ from fitting over $-2.0 \leq log_{10}\theta < -0.5$ using the combined field data for PE-\gzHKs\ (red dashed), SF-\gzHKs\ (blue dash-dotted), and all galaxies (gray dotted).
 }
\label{fig.12}
\end{figure}


\begin{table}
\centering
\caption{The estimated parameters from fitting the angular correlation function \omegals\ 
for the combined fields, with varying angular separation ranges. The parameter estimation method and the meanings of the quoted numbers are discussed in \S~\ref{sec.4.2}. The results vary depending on the fitting domain used; boldfaced values indicate our preferred fitting ranges and fit results. 
}
\begin{tabular}{@{}lcrc@{}}
\hline
Sample & $\log_{10}(\theta/1^\circ)$ & $\Aw / 10^{-2}$ & $\gamma$ \\
\hline

\sgzHK       & $[-2.0, -0.1)$ &  $5.09 _{-0.39}^{+0.37}$ & $1.73_{-0.05}^{+0.07}$ \\
\sgzHK 	    & $[-2.0, -0.5)$ &  $5.07 _{-0.40}^{+0.37}$ & $ 1.70_{-0.07}^{+0.08}$  \\
\sgzHK       & $[-2.0, -0.9)$ &  $5.05 _{-0.41}^{+0.38}$ & $1.61_{-0.08}^{+0.09}$ \\
\sgzHK       & $[-3.0, -0.1)$ &  $4.81 _{-0.27}^{+0.24}$ & $1.76_{-0.02}^{+0.03}$ \\
\sgzHK       & $[-3.0, -0.5)$ &  $4.80 _{-0.32}^{+0.27}$ & $1.76_{-0.02}^{+0.03}$ \\
\sgzHK       & $[-3.0, -0.9)$ &  $4.88 _{-0.29}^{+0.29}$ & $1.76_{-0.03}^{+0.03}$ \\
\hline
\pgzHK       & $[-2.0, -0.1)$ & $12.74_{-0.54}^{+0.53}$ & $1.90_{-0.05}^{+0.06}$ \\
\pgzHK       & $[-2.0, -0.5)$ & $12.76_{-0.54}^{+0.52}$ & $1.92_{-0.05}^{+0.06}$ \\
\pgzHK       & $[-2.0, -0.9)$ & $12.73_{-0.56}^{+0.52}$ & $1.87_{-0.06}^{+0.07}$ \\
\pgzHK       & $[-3.0, -0.1)$ & $12.98_{-0.46}^{+0.42}$ & $1.95_{-0.03}^{+0.03}$ \\
\pgzHK       & $[-3.0, -0.5)$ & $12.92_{-0.46}^{+0.42}$ & $1.96_{-0.03}^{+0.03}$ \\
\pgzHK       & $[-3.0, -0.9)$ & $13.01_{-0.47}^{+0.44}$ & $1.95_{-0.04}^{+0.03}$ \\
\hline
all galaxies & $[-2.0, -0.1)$ &  $2.83 _{-0.20}^{+0.18}$ & $1.60_{-0.05}^{+0.06}$ \\
all galaxies & $[-2.0, -0.5)$ &  $2.83 _{-0.20}^{+0.17}$ & $1.63_{-0.07}^{+0.09}$ \\
all galaxies & $[-2.0, -0.9)$ &  $2.84 _{-0.21}^{+0.17}$ & $1.60_{-0.09}^{+0.11}$ \\
all galaxies & $[-3.0, -0.1)$ &  $2.43 _{-0.14}^{+0.13}$ & $1.81_{-0.02}^{+0.03}$ \\
all galaxies & $[-3.0, -0.5)$ &  $2.33 _{-0.13}^{+0.13}$ & $1.83_{-0.02}^{+0.02}$ \\
all galaxies & $[-3.0, -0.9)$ &  $2.29 _{-0.12}^{+0.14}$ & $1.84_{-0.03}^{+0.02}$ \\
\hline

\end{tabular}
\label{tab:1}
\end{table}


In order to measure the large-scale clustering, fitting should be done over angular scales in which the contribution from the one-halo term is not significant, i.e., on scales larger than the break between the two slopes. From practical concerns, studies in the past have exercised varying degrees of care in this regard and consequently we fit over various ranges of angular separations to understand systematics, if any. Table~\ref{tab:1} lists the parameters fitted over various angular separation ranges. In all fields, $\omega_{LS}$ are sampled well up to $\log_{10} \theta = -0.1$ ($\theta$ in degrees). While we have the luxury of having four independent wide fields, the survey geometries vary significantly (Fig.~\ref{fig.1}), and the angular
correlation functions appear to reflect them at larger angular separations (\S~\ref{sec.4.4}).

In Fig.~\ref{fig.12}, we see that the estimated fitting parameters are indeed sensitive to the choice of fitting domain. This is particularly notable when the lower limit gets extended to smaller angular separation. However, so long as the lower limit remains above the one-halo/two-halo slope transition, the results remain consistent at $\sim$1$\sigma$ level, except for the narrowest interval of $-2.0 \leq \log_{10} \theta < -0.9$, for which the slopes $\gamma$ appear systematically underestimated. As the fitting domain gets extended to smaller angular separations, the power-law slope $\gamma$ becomes steeper for both PE-\gzHKs\ and SF-\gzHKs\ galaxies. This is likely a consequence of greater contribution to the correlation function from the one-halo term (Fig.~\ref{fig.11}).

\begin{figure}
\begin{center}
  \includegraphics[height=0.350\textheight]{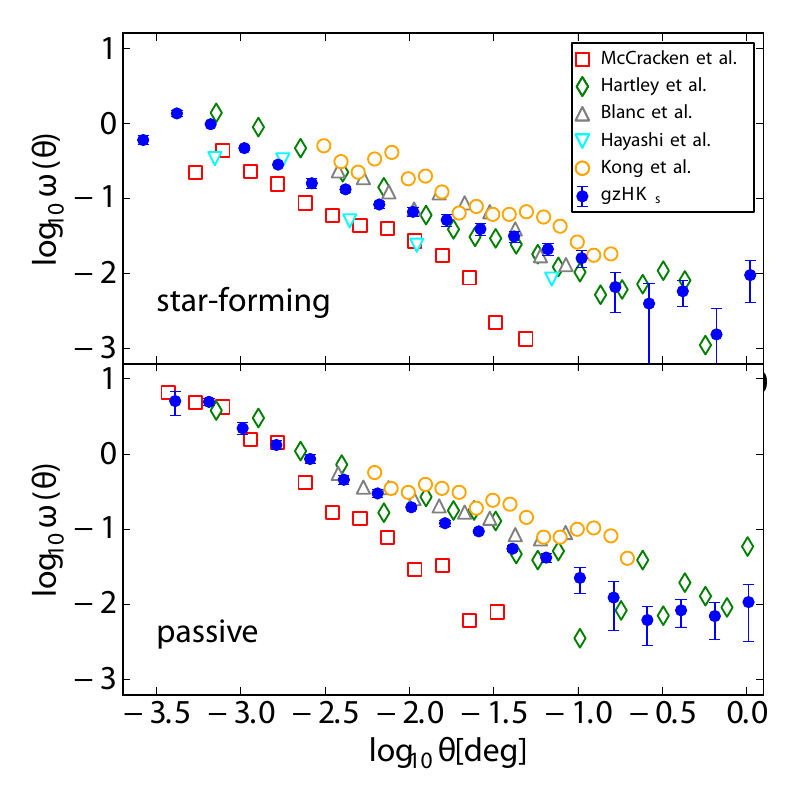}
\end{center}
 \caption[]{
 The angular correlation functions as a function of angular separation for SF-\gzHKs\ (upper panel) and PE-\gzHKs\ (lower panel) galaxies. The measurements are compared to other surveys employing similar color selection techniques (\BzK) that target $z\sim2$ galaxies.  For clarity the uncertainties for these other surveys' results are not plotted, but they are generally larger than those in our data. 
  }
\label{fig.13}
\end{figure}

Passive galaxies are known to preferentially reside in dense environments in the local universe (e.g., Dressler 1980; Balogh et al.\ 2004),  and it is likely that such an environmental relation was already present at $z\sim 1.5$ (e.g., Chuter et al.\ 2011; Quadri et al.\ 2012).  A stronger one-halo term contribution, in addition to the generally stronger clustering of PE-\gzHKs\ galaxies over the angular scales probed, could mean that a SFR-density relation similar to the one observed in the local universe already existed at $z\sim2$. While it is difficult to assess exactly how the clustering results relate to nearest-neighbor density estimates for local environment, the observation here suggests that the virialization of galaxies within massive halos has already progressed significantly by $z\sim2$ for both star-forming and passive galaxy populations, leading to situations in which environmental effects can be triggered to produce passive galaxies in such environments. In other words, the presence of the one-halo term for PE galaxies (in addition to that for SF ones) can be taken as evidence for environmental quenching at $z \sim 2$,  though it could also indicate other processes at play, such as mass quenching combined with a mass-density relation.

For the purpose of quantifying large-scale clustering, we wish to focus on scales where galaxy bias should be small (i.e., to isolate the two-halo term). In practice, the power-law fitting needs to be limited to above the break in the slopes of the angular correlation function and so we use the range of $-2.0 \leq \log_{10}\theta < -0.5$ for our clustering measurements unless otherwise noted. Strictly speaking, fitting above a certain angular range is not sufficient to isolate the clustering on the larger scales, as clearly seen in the difference between the single power-law fitting over $-2.0 \leq \log_{10}\theta < -0.5$ and the two-component power-law fitting over a larger angular scale in Fig.~\ref{fig.11}. In order to fit clustering observations better, it is best to cast the measurement in view of theoretical halo models, which we defer to a future opportunity. Nevertheless, the one-component analysis over a limited angular scale in this paper would still be useful to make comparisons to existing studies.


\begin{table*}
\centering
\caption{
Summary of selected recent \BzK-like clustering study surveys. The $K_{lim}$ are the magnitude cuts used for analysis. Note that the McCracken et al.\ (2010) field overlaps our D2 field. 
}
\begin{tabular}{@{}lccc@{}}
\hline
Study 					& Area [$deg^2$] 		& Subfields & $K_{lim}$ [AB] \\
\hline
Kong et al.\ (2006)		& 0.09 (Deep 3a-F) 		& 	2 (one of them is Deep 3a-F) 	& 	21.8 \\
Hayashi et al.\ (2007)	& 0.05				&	1						& 	23.2 \\	
Blanc et al.\ (2008) 		& 0.71				&	2						&	21.8 \\
Hartley et al.\ (2008) 	& 0.63				&	1						&	23 \\
McCracken et al.\ (2010)	& 1.9				&	1						&	23 \\
This work				& 2.59				&	4						&	23 \\
\hline

\end{tabular}
\label{tab:2}
\end{table*}

\begin{figure}
\begin{center}
  \includegraphics[height=0.320\textheight]{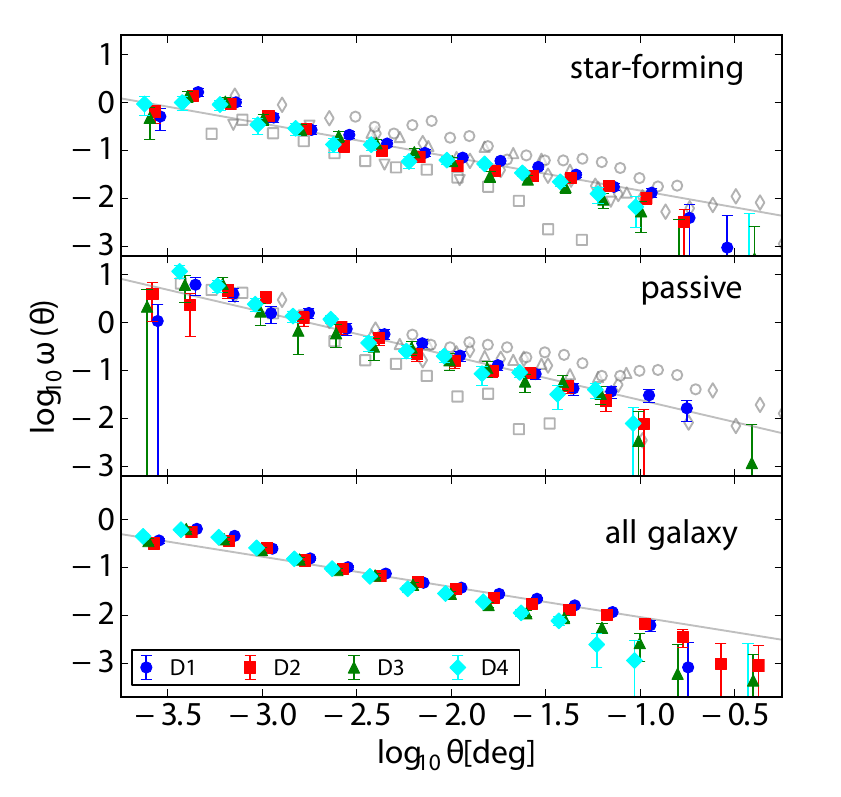}
\end{center}
 \caption[]{
 The angular correlation function $\omega_{LS}$ as a function of angular separation for SF-\gzHKs\ (upper panel) and PE-\gzHKs\ galaxies (middle panel), as well as all galaxies (lower panel) in each field. The errorbars indicate standard deviations computed in the standard manner within each angular separation bin; see \S~\ref{sec.4.2} for details. The solid gray lines are the best fitting power-law functions for the combined fields over the angular separation range of $-2.0 \leq log_{10}\theta < -0.5$ (\S~\ref{sec.4.3}). The various gray points in the upper and middle panels are the same results from the other surveys (Fig.~\ref{fig.13}) reproduced here for comparison.
 }
\label{fig.14}
\end{figure}

In Fig.~\ref{fig.13}, the combined angular correlation functions for all fields are compared to those
found in other studies (listed in Table~\ref{tab:2}). All these measurements use adaptations of the classic \bzk\ technique to select their samples, and therefore are directly comparable. For both SF-\gzHKs\ and PE-\gzHKs\ galaxies, the other correlation functions are roughly consistent, with ours being in the middle ground, although systematic differences obviously exist between different surveys. The correlation functions of Kong et al.\ (2006) are based on a shallower catalog ($K_{lim} \sim 21.8$) and sample a systematically more strongly clustered population than we do (\S~\ref{sec.4.5}). It is therefore not surprising that their values are systematically higher than ours. In contrast, while deep, the correlation functions of Hartley et al.\ (2008) are based on samples that do not accurately reproduce the classic \bzk\ selection (see Blanc et al.\ 2008; McCracken et al.\ 2010). Specifically, some of the galaxies that would be classified as passively evolving by classic \BzK\ selection (and in our PE-\gzHKs\ sample) are absent from the \BzK\ sample of Hartley et al.\ as they are lost predominantly out of the high-$z$ population into the low-$z$ galaxy selection window. While such loss affects the number counts and luminosity functions (see Arcila-Osejo \& Sawicki, 2013) it should not strongly affect clustering, unless very strong gradients are present in the clustering signal as a function of galaxy colour.  The fact that our clustering measurements agree with those of Hartley et al.\ (2008) confirms this view. Finally, especially curious are the correlation functions computed by McCracken et al.\ (2010) in the COSMOS field, which exhibit much steeper power-law behaviors than the rest of surveys shown in Fig.~\ref{fig.13}. Our D2 field is a subset of the COSMOS field, which is larger (2 deg$^2$ for COSMOS vs.\ 1 deg$^2$ for D2) so their pair counts must be very well sampled over the common angular separation intervals. The correlation functions of $gzHKs$ galaxies in the field computed in our study are not consistent with their steep power law functions (Fig.~\ref{fig.14}). The origin of this large discrepancy is unclear: although the COSMOS field is the most discrepant from the mean in terms of $z\sim 2$ galaxy number counts and luminosity functions (Arcila-Osejo \& Sawicki 2013), field-to-field cosmic variance (\S~\ref{sec.4.4}) is not sufficient to account for the differences in clustering between the McCracken et al. (2010) and our (and other) studies.

\subsection {Field-to-field Variations}\label{sec.4.4}

In Fig.~\ref{fig.14}, the angular correlation functions for SF-\gzHKs\ and PE-\gzHKs\ galaxies from all four CFHTLS deep fields are compared. The four fields are widely separated on the sky, so the sample variance (i.e., cosmic variance) is a significant contributor to the observed differences. The peculiar shapes of the survey geometry (Fig.~\ref{fig.1}) also lead to varying integral constraints. Such a sample variance is often conveniently cited for explaining apparent disparities among the results from various studies, yet even two closely related fields can yield quite different results based on separate analyses, as in the case for comparing $\omega_{LS}$ measurements for our D2 field to that of COSMOS (\S~\ref{sec.4.3}), Since each field in our study is as large as or larger than the typical survey area used in most of the past studies (Table~\ref{tab:2}), it would be beneficial to examine the effects of field-to-field variance when the same analysis is performed consistently on data from independent fields.

\begin{figure}
\begin{center}
  \includegraphics[height=0.320\textheight]{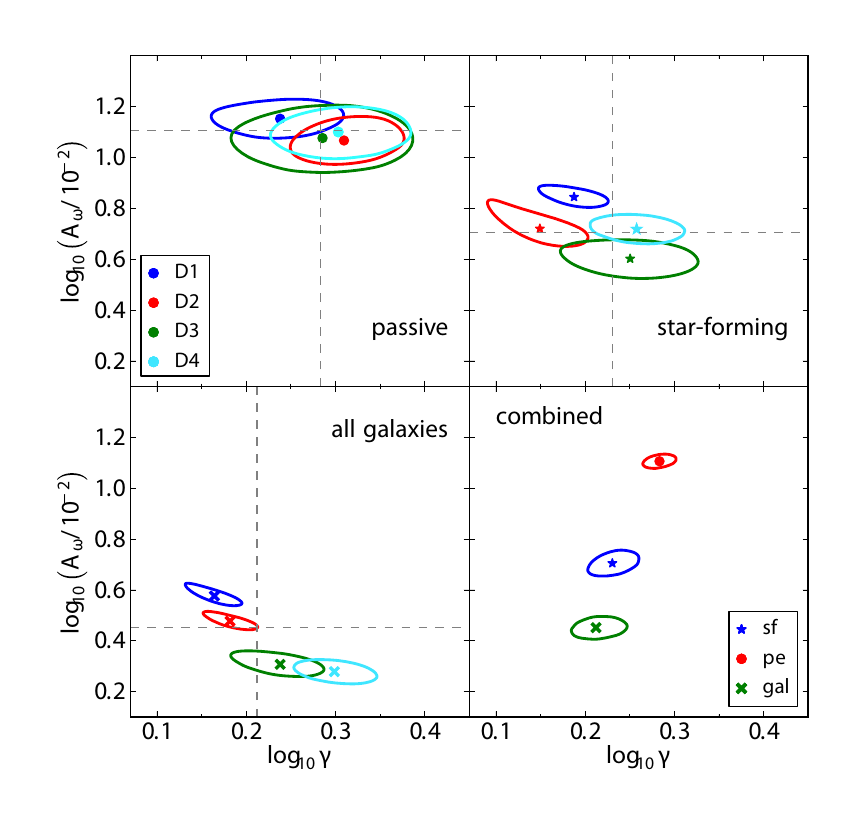}
\end{center}
 \caption[]{
 The field-to-field variations of the 68\% confidence intervals of the clustering amplitude \Aw\ and the power-law slope \g\ for PE-\gzHKs\ (upper left), SF-\gzHKs\ (upper right), all galaxies (lower left). The colors of contours indicate the fields D1 (blue), D2 (red), D3 (green), and D4 (orange). The results are also tabulated in Table~\ref{tab:3}. The black dashed lines indicate the minimum-\chisq\ parameters for each type of galaxies from the combined fields (Table 1), which are also shown in the lower right panel. The markers inside the contours indicate the parameter set for the minimum-\chisq\ fit. All results are from fitting over 
 $-2.0 \leq log_{10}\theta < -0.5$.
 }
\label{fig.15}
\end{figure}

Table~\ref{tab:3} shows the estimated parameters from fitting to $\omega_{LS}$ over $-2.0 \leq \log_{10}\theta < -0.5$  in the manner described in \S~\ref{sec.4.2}, allowing both the clustering amplitude $A_\omega$ and slope $\gamma$ free to vary (the integral constraint $C$ explicitly depends on $\gamma$).  Compared to the ÒgenericÓ values from the combined fields (Table~\ref{tab:1}), there exist significant field-to-field differences in the slope $\gamma$ of $\delta\gamma$ = 0.17, 0.12, and 0.21 for SF-\gzHKs, PE-\gzHKs, and all galaxies, respectively. The same information is graphically presented in Fig.~\ref{fig.15}, showing the field-to-field variation is at the 1--2$\sigma$ level. Since the clustering amplitude \Aw\ and slope $\gamma$ can be correlated (\S~\ref{sec.4.2}), they contain partially degenerate information about clustering strength. This makes it somewhat less intuitive when comparing measurements, especially given the amplitude $A_\omega$ is normalized in units of $[\theta]^{1-\gamma}$ and the authors vary on their choice of $\theta$ at which $A_\omega$ is defined as well as the value of $\gamma$ when fixed. It is best that \eq{eqn:6} be explicitly evaluated at a specific angular separation for comparison. Our clustering amplitudes $A_\omega$ are in units of arcminute$^{1-\gamma}$ with the standard deviations of $\sigma_{A_\omega} /10^{-2} = $1.06, 0.99, and 0.76 for SF-\gzHKs\, PE-\gzHKs\, and all galaxies, respectively.

\begin{table}
\centering
\caption{
The angular correlation functions $\omegals$ fitted separately for individual fields.
The estimated parameters are over the angular separation range between $-2.0 \le \log_{10}\theta < -0.5$, with all parameters allowed to vary.  The parameter estimation method and the meanings of the quoted numbers are discussed in 
\S~\ref{sec.4.2}.}

\begin{tabular}{@{}lcrcc@{}}
\hline
Sample & Field & $\Aw / 10^{-2}$ & $\gamma$ & $C / 10^{-1}$ \\
\hline
\sgzHK & D1 &  $6.99_{-0.38}^{+0.53}$ & $1.54_{-0.10}^{+0.08}$ & $1.94_{-0.41}^{+0.62}$ \\
\sgzHK & D2 &  $5.25_{-0.41}^{+1.09}$ & $1.41_{-0.15}^{+0.10}$ & $0.27_{-0.71}^{+1.56}$ \\
\sgzHK & D3 &  $4.00_{-0.42}^{+0.52}$ & $1.78_{-0.22}^{+0.20}$ & $1.17_{-0.44}^{+0.87}$ \\
\sgzHK & D4 &  $5.23_{-0.45}^{+0.46}$ & $1.81_{-0.14}^{+0.14}$ & $1.00_{-0.29}^{+0.44}$ \\
\hline
\pgzHK & D1 & $14.20_{-1.36}^{+1.73}$ & $1.73_{-0.20}^{+0.18}$ & $1.16_{-0.44}^{+0.88}$ \\
\pgzHK & D2 & $11.66_{-1.73}^{+1.46}$ & $2.04_{-0.18}^{+0.21}$ & $0.45_{-0.17}^{+0.26}$ \\
\pgzHK & D3 & $11.92_{-2.17}^{+2.17}$ & $1.93_{-0.30}^{+0.30}$ & $0.84_{-0.39}^{+0.90}$ \\
\pgzHK & D4 & $12.59_{-2.03}^{+1.65}$ & $2.01_{-0.22}^{+0.26}$ & $0.63_{-0.25}^{+0.42}$ \\
\hline
all galaxies & D1 & $3.77_{-0.19}^{+0.30}$ & $1.46_{-0.08}^{+0.06}$ & $2.46_{-0.40}^{+0.61}$ \\
all galaxies & D2 & $3.00_{-0.13}^{+0.18}$ & $1.52_{-0.07}^{+0.07}$ & $1.91_{-0.34}^{+0.47}$ \\
all galaxies & D3 & $2.03_{-0.13}^{+0.17}$ & $1.73_{-0.14}^{+0.13}$ & $1.35_{-0.36}^{+0.57}$ \\
all galaxies & D4 & $1.90_{-0.14}^{+0.13}$ & $1.99_{-0.14}^{+0.14}$ & $0.65_{-0.17}^{+0.24}$ \\
\hline
\end{tabular}
\label{tab:3}
\end{table}

\begin{table}
\centering
\caption{
The estimated parameters over the angular separation range $-2.0 \le \log_{10}\theta < -0.5$, with the power-law slope fixed at \g\ = 1.8. The amplitude \Aw\ is allowed to vary, and the integral constraint $C$ is computed via \eq{eqn:5}, i.e., effectively fixed for each field.}
\begin{tabular}{@{}lcrcc@{}}
\hline
Sample & Field & $\Aw / 10^{-2}$ & $C / 10^{-1}$ & $r_0$ [Mpc] \\
\hline
\sgzHK & D1 &  $6.53_{-0.33}^{+0.33}$ & $0.96$ & $7.87$ \\
\sgzHK & D2 &  $4.62_{-0.32}^{+0.32}$ & $0.86$ & $6.49$ \\
\sgzHK & D3 &  $3.99_{-0.42}^{+0.43}$ & $1.12$ & $5.98$ \\
\sgzHK & D4 &  $5.24_{-0.44}^{+0.44}$ & $1.03$ & $6.96$ \\
\hline
\pgzHK & D1 & $14.31_{-1.46}^{+1.47}$ & $0.96$ & $8.63$ \\
\pgzHK & D2 & $10.90_{-1.46}^{+1.47}$ & $0.86$ & $7.42$ \\
\pgzHK & D3 & $11.81_{-2.06}^{+2.06}$ & $1.13$ & $7.76$ \\
\pgzHK & D4 & $12.18_{-1.79}^{+1.79}$ & $1.03$ & $7.90$ \\
\hline
all galaxies & D1 & $3.23_{-0.10}^{+0.10}$ & $0.96$ & \\
all galaxies & D2 & $2.68_{-0.09}^{+0.08}$ & $0.86$ & \\
all galaxies & D3 & $1.99_{-0.12}^{+0.12}$ & $1.12$ & \\
all galaxies & D4 & $1.97_{-0.13}^{+0.13}$ & $1.03$ & \\
\hline

\end{tabular}
\label{tab:4}
\end{table}

\begin{table}
\centering
\caption{ The estimated parameters over the angular separation range $-2.0 \le \log_{10}\theta < -0.5$, with \g\ fixed at the values obtained from the combined fields.
  }
\begin{tabular}{@{}lcrccc@{}}
\hline
Sample & Field & $\Aw / 10^{-2}$ & $\gamma$ & $C / 10^{-1}$ & $r_0$ [Mpc] \\
\hline
\sgzHK & D1 &  $6.67_{-0.33}^{+0.33}$ & $1.70$ & $1.25$ & $8.84$ \\
\sgzHK & D2 &  $4.71_{-0.33}^{+0.32}$ & $1.70$ & $1.14$ & $7.20$ \\
\sgzHK & D3 &  $4.03_{-0.43}^{+0.43}$ & $1.70$ & $1.44$ & $6.57$ \\
\sgzHK & D4 &  $5.31_{-0.45}^{+0.45}$ & $1.70$ & $1.33$ & $7.73$ \\
\hline
\pgzHK & D1 & $14.49_{-1.51}^{+1.49}$ & $1.92$ & $0.70$ & $7.67$ \\
\pgzHK & D2 & $11.32_{-1.52}^{+1.51}$ & $1.92$ & $0.62$ & $6.74$ \\
\pgzHK & D3 & $11.92_{-2.04}^{+2.05}$ & $1.92$ & $0.85$ & $6.93$ \\
\pgzHK & D4 & $12.44_{-1.82}^{+1.80}$ & $1.92$ & $0.77$ & $7.08$ \\
\hline
all galaxies & D1 & $3.43_{-0.11}^{+0.11}$ & $1.63$ & $1.52$ & \\
all galaxies & D2 & $2.84_{-0.09}^{+0.09}$ & $1.63$ & $1.40$ & \\
all galaxies & D3 & $2.10_{-0.13}^{+0.13}$ & $1.63$ & $1.72$ & \\
all galaxies & D4 & $2.04_{-0.14}^{+0.14}$ & $1.63$ & $1.59$ & \\
\hline

\end{tabular}
\label{tab:5}
\end{table}

When observed angular correlation functions are noisy and of lower quality, a common practice is to fix the slope $\gamma$ at a canonical value such as $\gamma = 1.8$ which is based on low-$z$ measurements of galaxy clustering (e.g., Zehavi et al.\ 2002; Norberg et al.\ 2001), and simply let the clustering amplitude $A_\omega$ carry the clustering information. To facilitate comparisons, we present measurements with $\gamma = 1.8$ in Table~\ref{tab:4}. However, since the clustering amplitude $A_\omega$ can be sensitive to the slope $\gamma$ (Fig.~\ref{fig.12}) caution must be exercised when comparing clustering amplitudes measured with different values of $\gamma$. There is evidence that the power-law slope varies for different galaxy populations. For example, Adelberger et al.\ (2005) reports a flatter slope ($\approx 1.6$) for Lyman-break galaxies (i.e., high-$z$ star-forming galaxies), while McCracken et al.\ (2010) reports a much steeper slope ($\approx 2.5$) for passive \BzK\ galaxies. Our measurements do suggest a systematically steeper slope for PE-\gzHKs\, but the difference seems less pronounced than what these studies present (the McCracken et al.\ slopes are steeper than other studies for both passive and star-forming \BzK s). Few high-$z$ studies of this nature so far have allowed their correlation function slopes to vary, primarily due to data quality. Our work presents a significant improvement in the accuracy of the measurements over the previous studies.

\begin{figure}
\begin{center}
  \includegraphics[height=0.320\textheight]{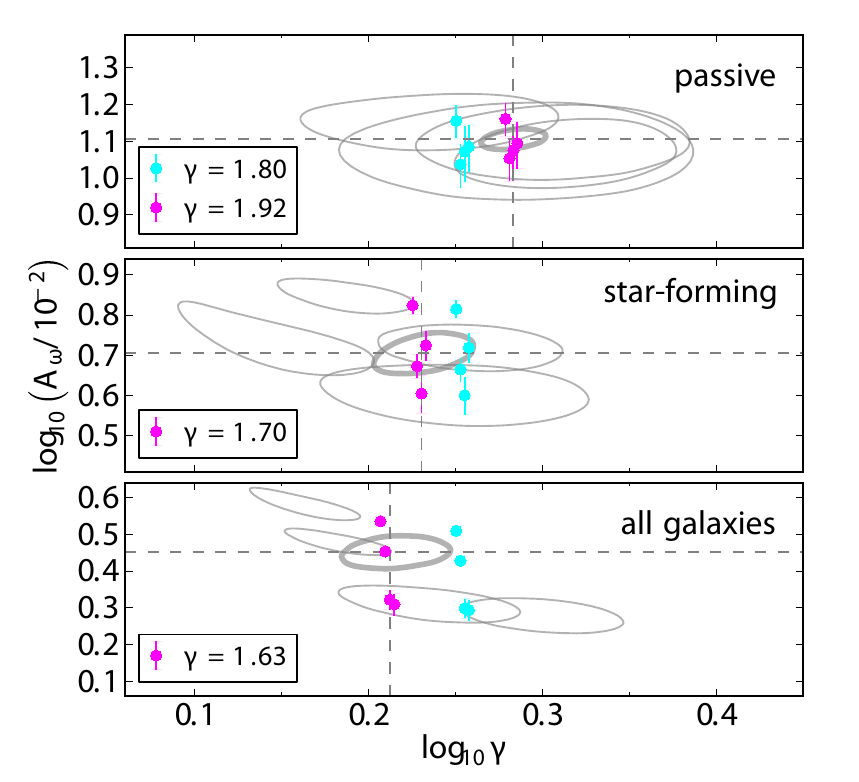}
\end{center}
 \caption[]{
The 68\% confidence contours of the clustering amplitude \Aw\ and the power-law slope \g\ for PE-\gzHKs\ (upper), SF-\gzHKs\ (middle), and all galaxies (lower). The thin gray contours are field-to-field results as in Fig.~\ref{fig.15}, shown for reference. The thicker gray contours are from the combined fields (Fig.~\ref{fig.12}). The cyan markers indicate the results when the power-law slope is fixed at \g\ = 1.8; the magenta markers are also for fixed slopes, \g\ = 1.92 for PE-\gzHKs, 1.70 for SF-\gzHKs, and 1.63 for all galaxies.  The plotted points are slightly offset horizontally for clarity. The errorbars for the fixed fitting do not exactly coincide with the 68\% confidence contours, since they are derived from independent MCMC sampling runs with fixed \g. All results are from fitting over 
  $-2.0 \leq log_{10} \theta < -0.5$.
 }
\label{fig.16}
\end{figure}

In Table~\ref{tab:5} we present the clustering amplitudes $A_\omega$ when the slopes are fixed at the values obtained from the combined fields: $\gamma = 1.70$ for SF-\gzHKs, 1.92 for PE-\gzHKs, and 1.63 for all galaxies. The same information is graphically presented in Fig.~\ref{fig.16}, in which we see that fixing the correlation function slope $\gamma$ does not lead to a confidence interval similar to the one obtained from marginal posterior distribution and therefore tends to underestimate the parameter uncertainty compared to the case when $\gamma$ is allowed to vary. On the other hand, so long as the amplitude $A_\omega$ and slope $\gamma$ are not strongly correlated (which appears to be the case for us), the estimated values for $A_\omega$ suffer little systematic effect by fixing $\gamma$. However, that is only true under a very fortuitous circumstance in which these parameters are not strongly correlated. We discussed in \S~\ref{sec.4.2}  that such a correlation apparently can become an issue depending on particular ways in which parameter estimation is carried out. To assess the potential for systematics introduced by fixed $\gamma$, it is suggested the \chisq\ space be inspected for correlation between $A_\omega$ and $\gamma$,  even if it is not explicitly used to estimate parameter uncertainty.

\begin{figure}
\begin{center}
  \includegraphics[height=0.290\textheight]{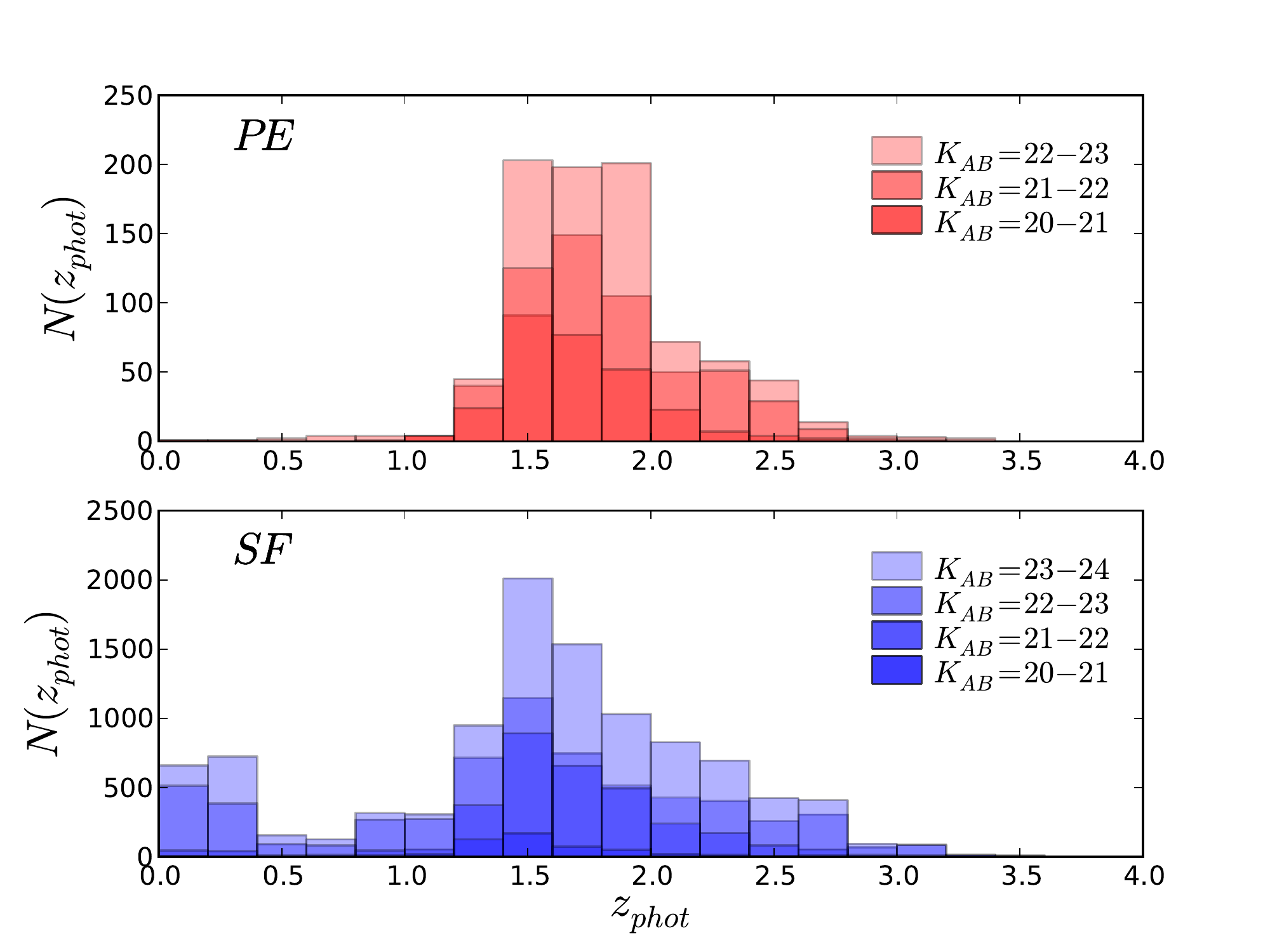}
\end{center}
 \caption[]{
Photometric redshift distributions of our PE and SF galaxies in the Muzzin et al.\ (2013) catalog.   
 }
\label{fig.17}
\end{figure}

It should be kept in mind that, for comparison purposes, the clustering amplitude \Aw\ is not a very meaningful indicator of clustering strength away from where ($\theta^{1-\gamma} - C$) is unity and when the power-law slope \g\ is allowed to vary. These are often degenerate, i.e., a large \g\ tends to be compensated by a small \Aw\ and vice versa. The same can be said for spatial correlation length $r_0$, which is a scale parameter in a power law of the form
\[
  \xi\left(r\right) = \left(\frac{r}{r_0}\right)^{-\gamma} \ ,
\]
and is estimated in a way that depends both on \Aw\ and \g\ via the Limber transformation (e.g., Limber 1953; Brainerd et al.\ 1995). In Table~\ref{tab:4} and Table~\ref{tab:5}, we list the values of $r_0$ computed via the same transformation, assuming Gaussian redshift distributions for star-forming and passive \BzK\ galaxies presented in Blanc et al.\ (2008)\footnote{The values are from Table~6 of Blanc et al.\ (2008): ($\bar z,\sigma_z) = (1.58, 0.16)$ and (1.78, 0.31) for passive and star-forming \BzK\ galaxies, respectively, where the $\sigma_z$ values reflect the intrinsic widths of the distributions after correction for photometric redshift scatter. We verified that the Blanc et al.\ redshift distributions are consistent with redshift distributions we computed for our PE and SF samples using the photometric redshift catalogs of Muzzin et al.\ (2013) in the D2/COSMOS field. We also find no evidence for strong differences in $N(z_{phot})$ as a function of magnitude within our samples (see Fig.~\ref{fig.17})}.

The correlation length $r_0$ is a measure of clustering scale, i.e., how galaxies' spatial distributions are biased, and compared among different surveys. In general, larger correlation lengths for the passive \BzK\ galaxies, compared to those for star-forming \BzK s, have been reported (Blanc et al.\ 2008; McCracken et al.\ 2010; Lin et al.\ 2012), which imply stronger clustering of passive galaxies. We confirm this tendency of PE-\gzHKs\ galaxies having generally larger correlation lengths than SF-\gzHKs\ galaxies, when the same slope \g\ is used for the two populations (Table~\ref{tab:4}).  When the correlation function slopes \g\ are not fixed at the same value, however, such a trend is actually not observed (Table~\ref{tab:5}) and in fact the correlation lengths for SF-\gzHKs\ are larger. Table 6 of McCracken et al (2010) shows a similar effect, although these authors do not explicitly comment on this issue. This exercise suggests that direct comparisons of correlation lengths may be relevant only under certain circumstances, such as when the slope \g\ is fixed for all populations being compared. As noted earlier, it has become apparent that the correlation function slope \g\ does vary between different galaxy populations at $z\sim 2$. Clearly, the full picture requires reliable measurements of both \Aw\ and \g, and caution should be exercised when interpreting correlation lengths computed with artificially fixed \g.

\subsection{Dependence on Galaxy Brightness}\label{sec.4.5}

The $K$-band, which at $z \sim 2$ samples the red part of the rest-frame optical/near-infrared SED, provides an often-used proxy for stellar mass. While a better mass estimate of a high-redshift galaxy can be obtained from full SED analysis (e.g., Sawicki \& Yee 1998), single-band estimates remain useful due to their simplicity.  Similarly, the rest-frame UV, at $z \sim 2$ sampled in the observed optical wavelengths, is a useful proxy for star formation activity in relatively unobscured galaxies (e.g., Kennicutt 1998; Sawicki 2012a).  Here we investigate the clustering of our \gzHK\ galaxies as a function of both observed $K$- and $r$-band magnitude. 

We note that artificial differences in clustering as a function of galaxy brightness can be introduced if, for example, there are differences between redshift selection windows for galaxies of different magnitude.  With that in mind, we checked for such N(z) differences within our samples by crossmatching the \gzHK\ galaxies in the D2 (COSMOS) field with objects in the Muzzin et al.\ (2013) photometric redshift catalog and we find that the N(z) distributions do not appear to vary strongly with object apparent magnitude.

\subsubsection{Rest-frame optical}\label{sec.4.5.1}

\begin{figure}
\begin{center}
  \includegraphics[height=0.350\textheight]{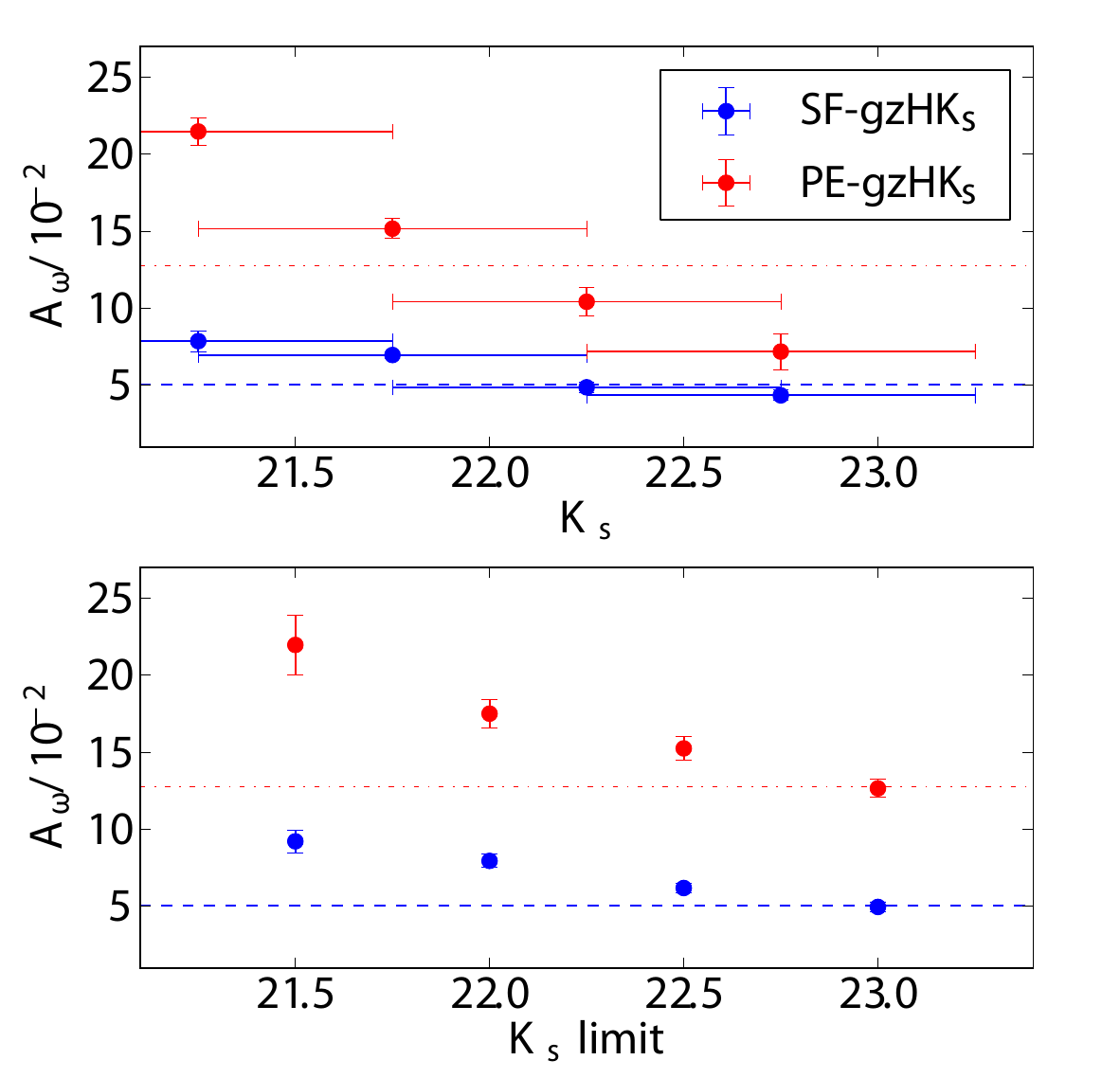}
\end{center}
 \caption[]{
  The clustering amplitude for SF-\gzHKs\ (blue) and PE-\gzHKs\ (red) galaxies as a function of $K_s$ magnitude. In the upper panel, the horizontal bars indicate the $K_s$ magnitude intervals defining subsamples. In the lower panel, the points are plotted at the upper limits for $K_s$ magnitude. All measurements are over $-2.0 < log_{10}\theta < -0.5$, with \g\ fixed at the values obtained for the combined field. The red dash-dotted and blue dashed lines indicate the values for the full samples.
 }
\label{fig.18}
\end{figure}

The relationship between $K$-band brightness and clustering strength is of interest because it relates to how galaxies of a given stellar mass cluster as a function of their dark matter halo mass.  

At $z = 0$, brighter galaxies exhibit stronger clustering (e.g., Zehavi et al.\ 2011). A qualitatively similar trend is observed for $K_s$-selected {\emph star-forming} \BzK\ galaxies at $z \sim 2$ (Kong et al.\ 2006; Hayashi et al.\ 2007; McCracken et al.\ 2010; Lin et al.\ 2012), which we confirm in Fig.~\ref{fig.18}. Additionally, we have a large enough sample to study the clustering of PE-\gzHKs\ galaxies as well, and their clustering trend is qualitatively similar to that of SF-\gzHKs s, though with a much stronger dependence on magnitude. This effect is not unexpected if PE-\gzHKs\ galaxies are indeed not forming stars:  if this is the case then they are dust-free and their observed $K$-band magnitudes correlate relatively straightforwardly with stellar mass. The newly-observed trend then indicates that more massive passive galaxies reside in more massive dark matter halos.

\begin{figure}
\begin{center}
  \includegraphics[height=0.355\textheight]{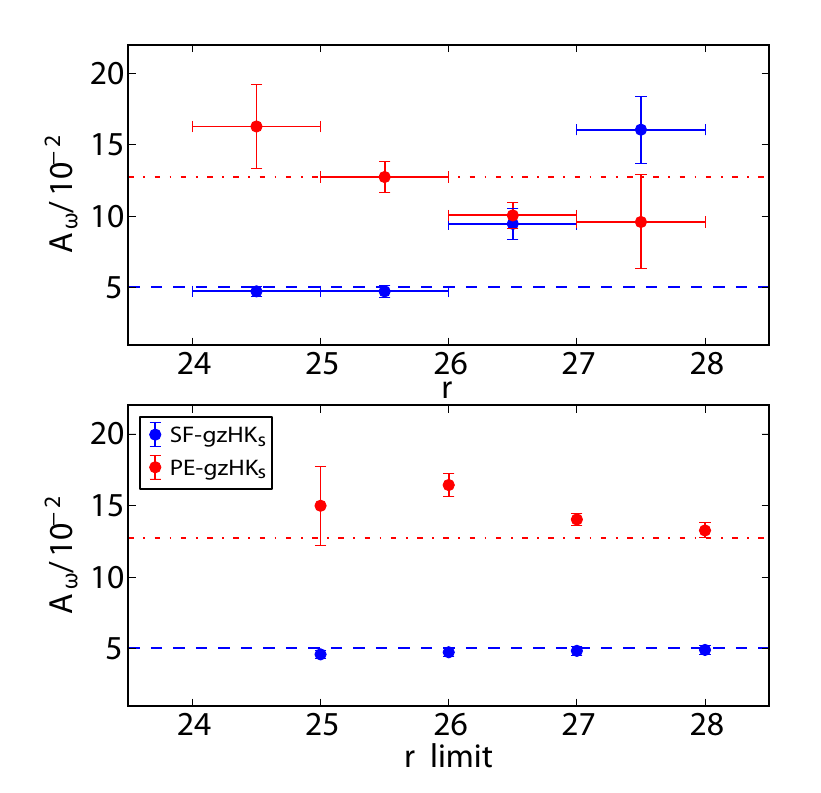}
\end{center}
 \caption[]{
The clustering amplitude for SF-\gzHKs\ (blue) and PE-\gzHKs\ (red) galaxies as a function of $r$ magnitude. In the upper panel, the horizontal bars indicate the $r$ magnitude intervals defining subsamples, not uncertainties. In the lower panel, the points are plotted at the upper limits for $r$ magnitude. All measurements are over $-2.0 < log_{10} \theta < -0.5$, with \g\ fixed at the values obtained for the combined field.
 }
\label{fig.19}
\end{figure}

\subsubsection{Rest-frame UV}

In star-forming galaxies (but not necessarily in passive ones), rest-frame UV luminosity traces the light from young stellar populations acting as a rough proxy for star-forming activity. In UV-selected samples at $z \sim 2$ and above, stronger clustering is observed for UV-bright galaxies, suggesting that the more vigorously star-forming galaxies are more clustered at those redshifts (e.g., Adelberger et al.\ 2005; Ouchi et al.\ 2004; Lee et al.\ 2006). At low redshift, $z \leq 1$, this trend with luminosity disappears or perhaps even reverses (Heinis et al. 2007).  At intermediate redshifts Savoy et al.\ (2011) report that the clustering as a function of UV luminosity (their observed $R$-band) turns over at $z \sim 2$ and reverses below that redshift; that is, at $z \ga 2$, UV-bright galaxies cluster more strongly compared to UV-faint galaxies, whereas at $z \la 2$ UV-faint galaxies cluster more strongly than UV-bright galaxies. If real, such a transition is very interesting, since the progression of the clustering peak presumably tracks the decrease of star formation in the most massive galaxies across the epochs at which cosmic star-forming activity peaks and starts to decline. 

Our SF-\gzHKs\ galaxies are at $z \sim 2$ or below (e.g., Blanc et al.\ 2008), so comparable to the BM ($z \sim 1.7$) galaxies in Savoy et al.\ (2011). In Fig.~\ref{fig.19}, we show the clustering amplitude as a function of $r$ magnitude for both SF-\gzHKs\ and PE-\gzHKs\ galaxies.  Our photometry is not as deep as that of Savoy et al.\ (2011), so it is difficult to asses how the results compare at the faintest magnitude bin, but even at brighter magnitudes interesting trends emerge.

A remarkable difference is observed in how SF-\gzHKs\ and PE-\gzHKs\ galaxies cluster in Fig.~\ref{fig.19} (top panel), where the clustering properties of \gzHKs\ galaxies within different r magnitude slices are plotted. For SF-\gzHKs, $r$-faint galaxies cluster more strongly --- in agreement with the BM (i.e., $z\sim 1.7$ star-formers) of Savoy et al. (2011) --- whereas the trend is opposite for PE-\gzHKs. Given that $r$-band samples the rest-frame UV, the behavior of SF-\gzHKs\ galaxies may indicate that those SF-\gzHKs s with less star-forming activity are found more clustered at that epoch already, consistent with lower-$z$ trends. A possible explanation of this trend is the vigorous activity seen in the most massive halos at earlier epochs is starting to shut down at $z \sim 2$ (Savoy et al.\ 2011). The case of PE-\gzHKs\ is more straightforward. They are selected for the lack of active star formation and their rest-frame UV light is produced by long-lived low-mass stars rather than recently-formed massive ones. Consequently, their luminosity is correlated with their stellar masses, and thus the clustering trend observed for PE-\gzHK\ galaxies likely simply reflects the fact that the more massive (hence rest-UV brighter) PE galaxies reside in the more massive DM halos, as we found in our analysis using the $K$-band (\S~\ref{sec.4.5.1}).

\subsection {Dependence on Color and Specific SFR}\label{sec.4.6}

\begin{figure}
\begin{center}
  \includegraphics[height=0.200\textheight]{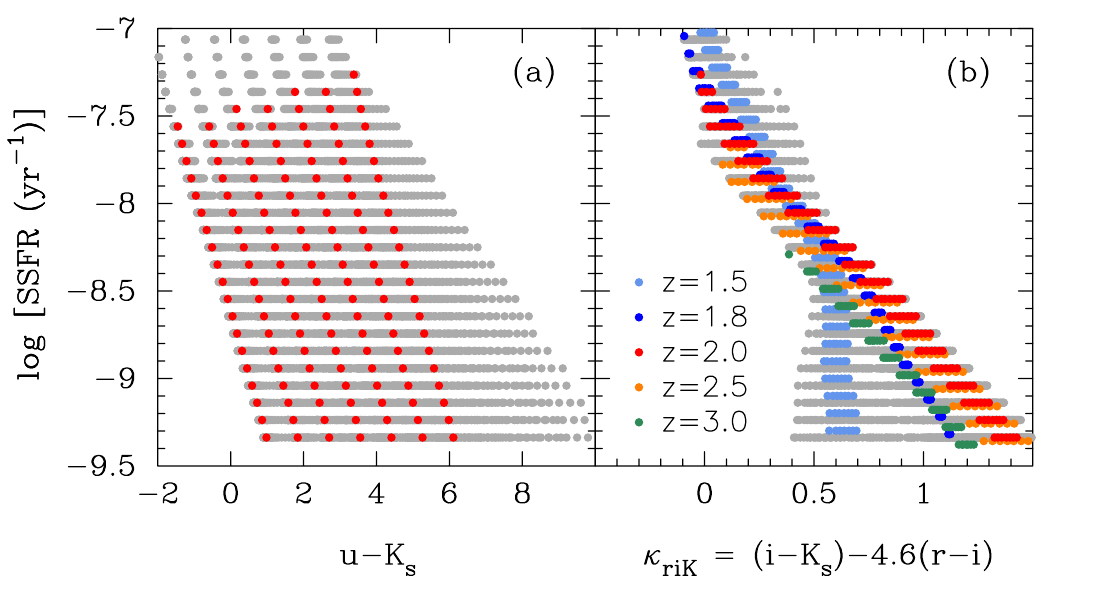}
\end{center}
 \caption[]{
 The specific SFR (sSFR) as a function of color combinations for SED models of SF-\gzHKs\ galaxies. Gray points show all the models considered, with a wide range of ages, redshift spanning $z = 1.5-3.0$, and $E(B-V) = 0.0-0.6$. Colored points highlight selected redshifts. The left panel shows $z = 2.0$ models only but with different amounts of extinction in order to highlight the fact that the use of a single color results in strong degeneracies in sSFR estimation. The right panel show a color combination that better correlates with sSFR. With the exception of objects at the very low end of the plausible redshift range, the color combination shown in the right panel identifies sSFR almost unambiguously, irrespective of reddening and across the likely redshift range spanned by SF-\gzHKs\ galaxies.
 }
\label{fig.20}
\end{figure}

\begin{figure}
\begin{center}
  \includegraphics[height=0.350\textheight]{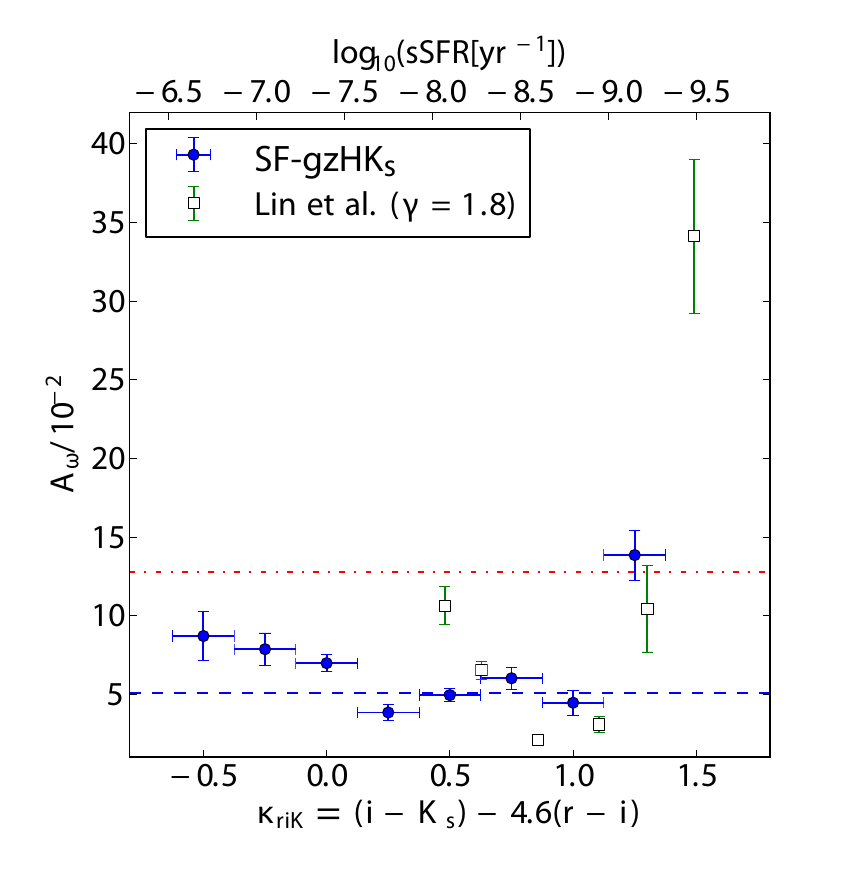}
\end{center}
 \caption[]{
 The clustering amplitude for SF-\gzHKs\ galaxies as a function of $\kappa_{riK}$, a proxy for sSFR (blue points with 1$\sigma$ error bars); see \S~\ref{sec.4.6} for the definition. The corresponding value of sSFR for $z \approx 2$ SF-\gzHKs\ galaxies is indicated on the upper axis. The color intervals are  $\kappa_{riK} = \pm0.125$ around the points shown. The blue and red dashed lines indicate the clustering amplitude \Aw\ of SF-\gzHKs\ and PE-\gzHKs\ galaxies for the combined fields, respectively. The open points are the measurements from Lin et al.\ (2012).
 }
\label{fig.21}
\end{figure}

There has been a recent suggestion by Lin et al.\ (2012) that the correlation length $r_0$ for star-forming \BzK\ galaxies exhibits a turnover as a function of specific star-formation rate (sSFR). That is, star-forming \BzK\ galaxies with low sSFRs are as strongly clustered as passive \BzK\ galaxies, but star-forming \BzK\ galaxies with high sSFRs are also more clustered than galaxies with more typical sSFRs; the minimum in the clustering strength appears to be at sSFR$\sim 2\times 10^{-9}$ yr$^{-1}$ (see Figure 5 in their paper). 

While we have not measured sSFR directly for our galaxies, a simple proxy measurements of sSFR can be obtained from carefully chosen combinations of broadband colors.  To do this, we determine the suitable colours using Bruzual \& Charlot (2003) spectral synthesis models adjusted for Calzetti et al. (1999) dust, cosmological effects, and integrated through filter transmission curves using SEDfit (Sawicki 2012b).  Naively, a reasonable choice for sSFR is $u-K_s$ color, since $u$ band traces the ultraviolet light from star formation while $K_s$ band is sensitive to the light from old stellar populations which constitutes the bulk of stellar mass in galaxies. However, dust reddening, which is also most severe in the rest-UV, dilutes the correlation between sSFR and $u-K_s$ color (Fig.~\ref{fig.20}(a)). Such a degeneracy can be alleviated to some extent by tracing $z\sim 2$ galaxy model tracks in a color-color plane in which dust degeneracy is projected out, effectively integrating over reddening variations. Such a transformation can be defined in the $riK_s$ color-color space: in Fig.~\ref{fig.20}(b), we see that the correlation between sSFRs and the lines of constant color combination are not as degenerate in terms of dust reddening as in $u-K_s$. Hence we define
\[
\kappa_{riK} \equiv (i-K_s) - 4.6(r-i)
\]
as our proxy for sSFR. In general, a higher value of $\kappa$ maps to a lower sSFR. We note that $\kappa_{riK}$ is a good proxy for sSFR at higher redshifts, $z\ga 1.7$, but at $z \sim 1.5$ starts to lose its ability to track sSFR at lower sSFR values  (light-blue curve in Fig.~\ref{fig.20}) and completely loses its ability to discriminate sSFR at lower redshifts, $z\la1.3$.  Consequently, if there is a large number of galaxies with $z \la 1.5$ in our sample then the strength of the $\kappa_{riK}$ sSFR proxy will be diluted (but not erased) by mixing galaxies where $\kappa_{riK}$ has discriminating power with "noise" from those without it. With these caveats in mind, the clustering as a function of sSFR is summarized in Fig.~\ref{fig.21}.

In both Lin et al.\ (2012) and our SF-\gzHKs\ sample, it is clearly evident that the most quiescent $z \sim 2$ star-forming galaxies are strongly clustered compared to the rest. This is another confirmation that quiescent galaxies are generally more strongly clustered than active galaxies. Lin et al.\ (2012) additionally reported that their star-forming \BzK\ galaxies with high sSFR (hence low $\kappa$ in our analysis) were more clustered than those with moderate sSFR, the correlation length $r_0$ effectively exhibiting a turnover as a function of sSFR. In Fig.~\ref{fig.21}, we do not see the clustering amplitude exhibiting such a drastic turnover in our sample, although there may be a slight, gradual increase in clustering amplitude toward higher sSFR. However, the comparison is complicated by the fact that the sSFRs are derived by quite different methods. The crude nature of transformation between $\kappa$ and sSFR here should be remembered, especially away from $z = 2$ where the degeneracy among different model tracks can become severe; as mentioned in the previous section, the bulk of SF-\gzHKs\ galaxies can be expected to be at $z < 2$ if the redshift distribution of the SF-\gzHKs\ galaxies were to follow those presented in previous studies. The issue of clustering dependence on sSFR is worth revisiting in the future with more precise sSFRs measurements.

\section{Conclusions}\label{sec.5}

We presented the angular clustering analysis of $z \sim 2$ galaxies selected via the \gzHKs\ color-color method, a substitute for the popular \BzK\ technique that we adapted for the CFHT observations of the four CFHTLS deep fields. The angular correlation functions were computed via the Landy-Szalay estimator, both for each field independently and for the combined fields. With four independent fields covering in total $\sim$2.5~deg$^2$, we have one of the largest-area deep surveys and could constrain $z\sim 2$ galaxy clustering through the unprecedented amount of data.  In summary, our main results are:

\begin{enumerate}
\item Over the angular scale of $-2.0 \leq \log_{10} \theta < -0.5$, the fitted power-law slopes \g\ are 
$1.74^{+0.21}_{-0.21}$ and $1.91^{+0.34}_{-0.30}$ for SF-\gzHKs, PE-\gzHKs\ galaxies, respectively. The corresponding clustering amplitudes (in units of 10$^{-2}$ arcmin$^{-(1-\gamma)}$) are $5.15^{+0.81}_{-0.80}$ and $12.73^{+2.45}_{-2.77}$ for the same galaxy types. We thus robustly confirm previous results that passive galaxies are more clustered (i.e., steeper slope $\gamma$ as well as larger angular clustering  amplitude \Aw) than star-forming galaxies at $z\sim 2$.

\item We find evidence for the existence of two components in angular correlation functions not just for SF galaxies (as was previously known) but also for PE galaxies.  We attribute these 
to the longer-scale clustering of underlying dark matter halos (the two-halo term) and multiple galaxies associated with individual halos (the one-halo term). 
The existence of a one-halo terms for both PE and SF galaxies suggests that  the virialization of galaxies within massive halos has already progressed significantly by $z \sim 2$ for both star-forming and passive galaxy populations, leading to situations in which environmental effects can be triggered to produce passive galaxies in such environments. In other words, the presence of the one-halo term for PE galaxies (in combination with that for SF ones) can be taken as evidence for environmental quenching at $z \sim 2$. The break between the one- and two-halo terms appears to be at the angular scale $\log_{10}\theta \approx -2.7$ for both populations, corresponding to 60 kpc at $z \sim 2$. 

\item We find that the clustering strength of PE galaxies increases as a function of rest-frame UV luminosity, indicating that the more massive among these passive galaxies reside in more massive DM halos. In dramatic contrast, the clustering of SF galaxies decreases with increasing UV luminosity. A possible explanation of this trend is that the vigorous star-forming activity seen at earlier redshifts ($z \ga 3$) in the most massive halos is starting to shut down around \zs2 leaving the most massive (and thus most clustered) halos with only low levels of star formation and, thus, low UV luminosities (see Savoy et al.\ 2011).

\item There are noteworthy differences (at the 1--2$\sigma$ level) in the best-fit clustering parameters between our four large, widely-separated fields. Notably, the difference from the combined-fields value is most pronounced for the D2 field, which is overlapped by the popular COSMOS field. The number counts of \zs2 galaxies in D2/COSMOS are also anomalous compared to the other fields (Arcila-Osejo \& Sawicki 2013), and together these results highlight the need to study several spatially-independent areas as is done in the present work.

\end{enumerate}

A natural continuation of the two-dimensional clustering analysis presented in this paper is to extend the analysis to three-dimensional clustering. Doing so would allow us to better link the observed clustering to the masses of the underlying dark matter halos, and --- in conjunction with stellar mass estimates --- to gauge the connection between stars that are forming, stars that already formed, and the dark matter halos that their host galaxies reside in.  We hope to do so at the next opportunity.

\section*{Acknowledgments}

We thank ACEnet and its staff for providing us with a wonderful high-performance computing environment without which the research presented here would not have been possible. We thank Sergiy Khan for his professional assistances throughout the project and for his desire to provide users with the best computing experiences. We thank Andrew Becker for making HOTPanTS publicly available, as well as for numerous assistances in getting it to run properly and understanding the underlying principle. We thank Jerzy Sawicki for a careful reading of the manuscript and the referee for comments that helped improve the quality of this paper. This work benefited tremendously from the Python programming language, its tools, and the community.  Computational facilities for this work were provided by ACEnet, the regional high performance computing consortium for universities in Atlantic Canada. ACEnet is funded by the Canada Foundation for Innovation (CFI), the Atlantic Canada Opportunities Agency (ACOA), and the provinces of Newfoundland and Labrador, Nova Scotia, and New Brunswick. This research was financially supported by a Discovery Grant from the Natural Sciences and Engineering Research Council of Canada (NSERC) and by an ACEnet Fellowship.

This work is based on
    observations obtained with MegaPrime/MegaCam and WIRCam.  The
    former is a joint project of CFHT and CEA/DAPNIA, at the
    Canada-France-Hawaii Telescope (CFHT) which is operated by the
    National Research Council (NRC) of Canada, the Institut National
    des Science de l'Univers of the Centre National de la Recherche
    Scientifique (CNRS) of France, and the University of Hawaii.  The
    latter is a joint project of CFHT, Taiwan, Korea, Canada, France,
    at the CFHT which is operated by the NRC of Canada, the Institute
    National des Sciences de l'Univers of the Centre National de la
    Recherche Scientifique of France, and the University of Hawaii.
    This work is based in part on data products produced at TERAPIX
    and the Canadian Astronomy Data Centre as part of the
    Canada-France-Hawaii Telescope Legacy Survey, a collaborative
    project of NRC and CNRS.  This work is also based in part on data
    products produced at TERAPIX, the WIRDS (WIRcam Deep Survey)
    consortium, and the Canadian Astronomy Data Centre. This research
    was supported by a grant from the Agence Nationale de la Recherche
    ANR-07-BLAN-0228

\appendix 

\section{Producing PSF-matched Images}
\label{sec:appendixA}

Substantial PSF variations exist across the individual stacked
images.  In visible bands, such PSF variations are typically smooth
over the entire image and can be modeled by low-order polynomials.  In
near-infrared images, variations are much more complicated due to
dithering and mosaicing the limited field of view
(\fig{fig.1}).  In order to better sample the light from the
same parts of objects, images are generally degraded to the worst
seeing before carrying out fixed-aperture photometry.

A customized version of
HOTPanTS\footnote{http://www.astro.washington.edu/users/becker/hotpants.html}
by Andrew Becker is used to find convolution kernels that match PSFs
between two images.  The program implements the algorithm devised by
\citet{alar00}, constructing a spatially-varying kernel $K$ that
minimizes the differences between point-source stamps in reference and
given images $R(x, y)$ and $I(x, y)$, respectively:
\begin{equation}
  \sum_i \left( [R \otimes K](x_i, y_i) - I(x_i, y_i) \right)^2 \ .
  \label{eqn:alard}
\end{equation}
The kernel $K$ is decomposed into some basis functions
\[
  K(u, v) = \sum_n a_n K_n(u, v)
\]
which should be any reasonable orthogonal functions but typically
Gaussians with varying widths are used:
\[
  K_n(u, v) = e^{-(u^2 + v^2)/2\sigma_k^2} u^i v^j
\]
where $n = (i, j, k)$.  The solutions found over the image are
connected via
\[
  K(u, v) = \sum_n a_n(x, y) K_n(u, v)
\]
where
\[
  a_n(x, y) = \sum_{i, j} b_{i,j} x^i y^j \ .
\]
This effectively finds a polynomial solution that models the
smoothly-varying convolution kernel over the entire (or parts of an)
image.  The primary advantage of the algorithm is that it makes
computation of spatially-varying kernel efficient; see Alard (2000)
for details.

\begin{figure*}
\includegraphics[width=16cm]{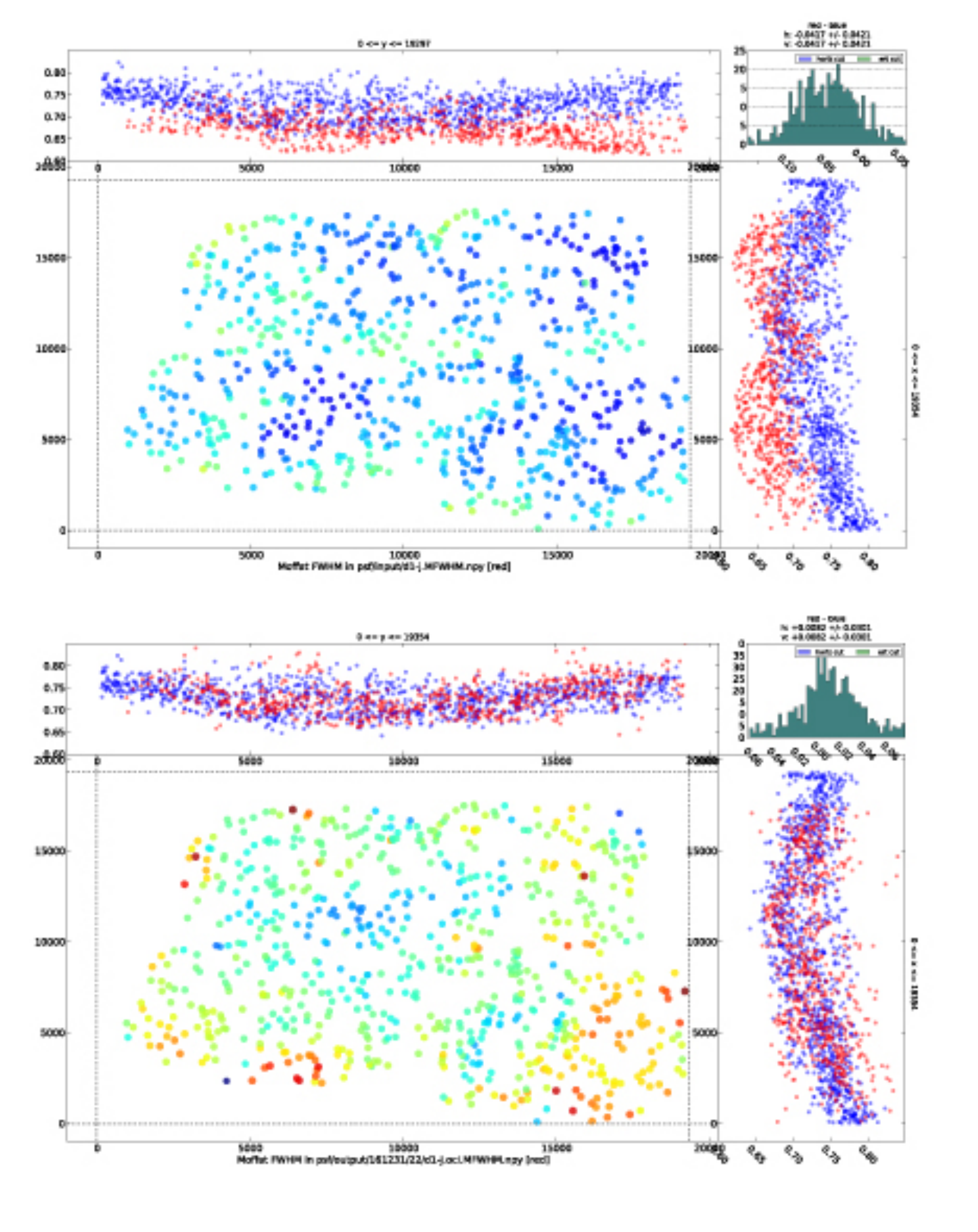}
  \caption{The output from the graphical user interface for the PSF matching utility (between $u$ and $J$ bands in the D1 field in this specific case). The top panel is for before PSF matching and the bottom panel after. The main plot shows the spatial distribution of point sources used for PSF matching, with the colors mapped to their FWHMs. The side plots indicate the spatial variation of the point source size in $u$-band (blue) and $J$ band (red). The spatial variations look worse than they actually are, since the data points are collapsed both vertically and horizontally over the whole image. In practice smaller slices are made for inspection which make the scatters after the PSF matching more reasonable.}
  \label{fig:psfmatch_1}
\end{figure*}

Originally designed for detecting time-varying signals by subtracting
one PSF-matched image from another, HOTPanTS implementation assumes
that images $R$ and $I$ are taken through the same passband, with the
only variable being their PSFs (due to different seeing, for example).
Our modifications are made to match PSFs on images taken through
different passbands.  This involves an introduction of a factor $r_i$
in \eq{eqn:alard} which takes into account the flux ratios of point
sources in $R$ and $I$ that are used to find the kernel $K$:
\begin{equation}
  \sum_i \left( [R \otimes K](x_i, y_i) - r_i I(x_i, y_i) \right)^2 \ .
\end{equation}
SExtractor is run on all images individually to construct point
source catalogs (via CLASS\_STAR), which are candidate point sources to be used in PSF
matching (\fig{fig:psfmatch_1}).  Their FWHMs and light profiles are
visually inspected and those sources with anormal values are removed
from the subsequent processing.  The modified HOTPanTS is then used to
match PSFs using the point sources and their flux ratios between two
images (\fig{fig:psfmatch_1}).

Since the kernel basis functions that we use are fairly generic
superpositions of three or four Gaussians with varying $\sigma_k$, our
ability to accurately match PSFs between two images is limited by the
rather arbitrary choice of kernel basis.  In practice, however, most
existing studies use a simpler convolution kernel for PSF matching on
a smaller field, if at all.  The degree of desired accuracy really
depends on science goals, and given our wide field of view and the
significant differences in seeing/image quality through different
passbands, the PSF correction at the level we carry out is necessary
and sufficient.  For fields D1, D3, and D4, the $u$ band image has the
largest PSFs ($\approx 0.75''$), to which the rest of the bands are
convolved.  In D2, the $J$ band has the largest PSFs ($\approx 0.9''$)
and the other images were convolved to that band.

\bsp

\label{lastpage}

\end{document}